\documentclass[aps,physrev,twocolumn,showpacs,superscriptaddress,amsmath,amssymb,amsfonts,floatfix]{revtex4-2}

\usepackage[utf8]{inputenc}
\usepackage[T1]{fontenc}        % Important for anything that uses more than ASCII
\usepackage{lmodern}            % Beautiful fonts especially for pdfs
\usepackage{graphicx}
\usepackage{multirow}
\usepackage{epsfig}
\usepackage{url}
\usepackage[normalem]{ulem} % [normalem] makes et al. to be italic instead of underlined in references
\usepackage[colorlinks,breaklinks,bookmarks=true,citecolor=blue,linkcolor=blue,urlcolor=blue]{hyperref}
\usepackage{tabularx}
\usepackage{sidecap}
\newcommand{\orcidlink}[1]{}

\usepackage{amsmath}            % Better and more beautiful mathematical formulas
\usepackage{amstext}            % \text{} macro in math mode
\usepackage{amssymb}            % Special mathematical symbols
\usepackage{mathtools}          % Even more beautiful formulas
\usepackage{bm}                 % bm
\usepackage{scalerel,stackengine} % stacking math expressions
\usepackage{ulem}
\usepackage{xcolor}

\begin{document}

\title{Orbital character and photoemission signatures of the $A$-pocket in infinite-layer nickelates}

\author{Leonard M. Verhoff\orcidlink{0009-0004-3358-1312}}
\affiliation{Institute of Solid State Physics, TU Wien, 1040 Vienna, Austria}

\author{Michael Seidl\orcidlink{0009-0005-2638-7551}}
\affiliation{Institute of Solid State Physics, TU Wien, 1040 Vienna, Austria}

\author{Liang Si\orcidlink{0000-0003-4709-6882}}
\affiliation{School of Physics, Northwest University, Xi'an 710127, China}
\affiliation{Institute of Solid State Physics, TU Wien, 1040 Vienna, Austria}

\author{Karsten Held\orcidlink{0000-0001-5984-8549}}
%\thanks{Corresponding authors:}
\email[]{Corresponding authors: leonard.verhoff@tuwien.ac.at; held@ifp.tuwien.ac.at}
\affiliation{Institute of Solid State Physics, TU Wien, 1040 Vienna, Austria}

\begin{abstract}
The orbital character of the so-called $A$-pocket 
%momentum \khx{electron} 
 in infinite-layer nickelates remains under debate. We investigate bulk NdNiO$_2$ using density-functional theory combined with dynamical mean-field theory and identify the $A$-pocket as a strongly hybridized state with substantial interstitial-$s$, Nd-$d_{xy}$ and Ni-$d_{xz/yz}$ contributions. Our angle-resolved photoemission spectroscopy simulations reproduce the main polarization-dependent features measured experimentally. Removing the interstitial-$s$ matrix element leads to a clear disagreement with the measured Fermi surface maps,
even though the interstitial-$s$ contribution is not dominant. Suppressing instead the Nd-$d_{xy}$ or Ni-$d_{xz/yz}$ matrix elements leaves the maps almost unchanged, because these are small at the zone corner.
More generally, our results demonstrate that assigning the microscopic origin of spectral features in infinite-layer nickelates requires treating orbital hybridizations and photoemission matrix elements together.
%More generally, our results demonstrate that orbital hybridization and photoemission matrix elements must be treated together when assigning the microscopic origin of spectral features in infinite-layer nickelates.
\end{abstract}

\date{\today}

\maketitle

% \kh{ToDo:\\
% \begin{itemize}
%     \item $R$-point $\rightarrow$ R-point?
%     \item In the SM there is a bit of question how to present Bloch vs. Wannier. First though the presentation was for Wannier. In the Bloch basis there also would not be off-diagonal elements for DFT. For DMFT we cannot make a Bloch basis for all frequencies with the same rotation.
    
% \end{itemize}
% }

\section{Introduction}
The arguably most striking difference between cuprate~\cite{Bednorz1986} and infinite-layer nickelate superconductors~\cite{li2019superconductivity} is the electron pocket at the $A$-point which, in addition to the Ni-$d_{x^2-y^2}$ band, crosses the Fermi energy.
This $A$-pocket, along with a $\Gamma$-pocket,
was reported in density-functional theory (DFT) calculations \cite{Nomura2019,PhysRevLett.125.077003}.
Including electronic correlations, e.g., within dynamical mean-field theory (DMFT),
shifts both pockets up in energy \cite{Si2020}. Although correlations strongly
shift the $\Gamma$-pocket for LaNiO$_2$ even above the Fermi energy, the $A$-pocket nevertheless remains at the Fermi energy, even when accounting for electronic correlations and hole doping.

Experimentally, a first indication for the presence of the $A$-pocket was the negative Hall resistivity \cite{li2019superconductivity,zeng2020}
in a wide range of doping concentration and temperature.
More direct evidence was then provided by angle-resolved photoemission spectroscopy (ARPES) for LaNiO$_2$ \cite{Ding2024,Sun2025} and NdNiO$_2$ \cite{li2025}, clearly
resolving an $A$-pocket but no $\Gamma$-pocket. A comparison of ARPES with DFT+DMFT spectral functions for LaNiO$_2$ can be found in \cite{Si2024}; surface effects are discussed in \cite{Verhoff2025}.

Li \emph{et al.}~combined polarization-dependent ARPES with resonant photoemission to determine the orbital character of the low-energy states \cite{li2025}.
%Based on polarization selection rules and a comparison with 
Based on atomic-orbital matrix elements calculated in Ref.~\cite{Ye_2013}, they concluded that the $A$-pocket is of interstitial-$s$ character, as proposed theoretically in Refs.~\cite{Foyevtsova2023,Gu2020}.
%
%disfavored dominant Nd-$d_{xy}$ and Ni-$d_{xz/yz}$ contributions to the $A$-pocket and assigned it predominantly to electride-like interstitial-$s$ states, as proposed theoretically in Refs.~\cite{Foyevtsova2023,Gu2020}.
Resonant photoemission across the Nd $4d\rightarrow4f$ and $3d_{5/2}\rightarrow4f$ absorption edges revealed Nd-$4f$-related features at larger binding energies, but no detectable resonant enhancement within 1\,eV of the Fermi level.
This indicates that localized Nd-4$f$ states contribute negligibly to the low-energy bands and Fermi surface, in line with resonant photoemission on PrNiO$_2$, where rare-earth $4f$ states were found to hybridize only deeper in the valence band while the spectral weight at $E_F$ was attributed to rare-earth $5d$ states~\cite{Chen2022Matter}.
Determining the orbital nature of the $A$-pocket and understanding the polarization-dependent ARPES experiments motivate our work.

Using DFT and DMFT, we find that the $A$-pocket at the Fermi energy is of mixed Ni-$d_{xz/yz}$, interstitial-$s$, and Nd-$d_{xy}$ character.
By calculating polarization-dependent ARPES spectra within the widely employed three-step model of photoemission \cite{MOSER2017,PhysRevX.12.011019,g9d4-qls9,Yen2024,Beaulieu2020,Day2019}, using a plane-wave final-state approximation (PWA) \cite{Williams1977,Puschnig2009}, we find that this mixed character is consistent with the principal polarization-dependent features observed experimentally.
Our simulations combine PWA matrix elements \cite{Yen2024,MOSER2017} with the full spectral-function orbital matrix, so that the photocurrent is evaluated as a coherent sum---rather than as a weighted trace---that includes interorbital interference. Although more sophisticated descriptions of the final states exist within the three-step model \cite{Yen2024,Kern2023,Ryoo2025} and the one-step model of photoemission \cite{Braun2018,Minar2011,Minar2020}, PWA keeps the orbital-resolved matrix elements transparent and permits controlled tests in which selected matrix-element channels are suppressed.

The remainder of this paper is organized as follows. Section~\ref{Sec:Comp} presents the computational methods and the framework used to simulate ARPES intensities. In Sec.~\ref{Sec:results}, we first analyze the hybridization and orbital character of the $A$-pocket and then compare the simulated polarization-dependent ARPES maps with experiment. Section~\ref{Sec:conclusion} summarizes our findings and their implications for assigning the orbital character of the $A$-pocket. Additional methodological details and results are provided in the Supplemental Material \cite{SM}.

% \khc{Not sure what to keep of the following Intro part/sketch. Maybe rather keep it for the
% method paper?}

% \lv{The process of decomposing ARPES signal into orbital contributions is typically named Photoemission orbital tomography.\\
% Plane-wave final state approximation is a crude approximation that covers the orbital symmetries, however, it neglects final-state multiple scattering as well as surface and layer interference arising from the correct boundary conditions. This is distinct from the interorbital interference retained through the coherent sum in the photocurrent. Therefore, it might work for qualitative comparison to ARPES images, however, for subtle effects such as circular dichroism it horribly fails. For that one typically employs the full treatment in one step model [Jan], or even three step model approaches with elaborate description of final states [Schüler]. Recent development of formalism for TRLEED states in plane-wave DFT [Korean guys]\\
% However, one advantage of modeling in 3-step model is, that we can directly separate the signal into contributions coming from different orbitals.\\}

\section{Methodology {and} Computational Details}
\label{Sec:Comp}
Similar to previous theoretical studies on NdNiO$_2$ \cite{Si2020,Kitatani2020,DiCataldo2023b,Verhoff2025}, we performed electronic structure calculations with $P4/mmm$ (No.~123) symmetry and $a$=$b$=3.854\,\AA, $c$=3.265\,\AA.
The Perdew--Burke--Ernzerhof version of the generalized gradient approximation (GGA-PBE) was employed to treat the exchange-correlations \cite{Perdew96}, as implemented in the all-electron full-potential code \textsc{WIEN2K} \cite{wien2k}.
Since (non-magnetic) DFT calculations incorrectly place the Nd-4$f$ states at the Fermi level, the open-core approximation was adopted to treat the Nd-4$f$ orbitals as core states.
We downfold the full space of Bloch states to a low-energy 11-band tight-binding model (including Nd-5$d$+Ni-3$d$+interstitial-$s$) with maximally localized Wannier functions calculated via \textsc{WANNIER90} \cite{Pizzi2020} interfaced to \textsc{WIEN2K} by \textsc{WIEN2WANNIER} \cite{Kunes2010a}.
The considered 11-band tight-binding model consists of five Ni-3$d$-derived and five Nd-$5d$-derived states, supplemented by an interstitial-$s$ state. The interstitial-$s$ state is located in the interstitial region between two Ni sites, where apical oxygen would sit in the stable, perovskite parent phase (see SM \cite{SM} Sec.~\ref{Sec:SM_wannierization} for the DFT band structure compared to the Wannier model and an illustration of interstitial-$s$ orbital).

Subsequently, a local Hubbard--Kanamori interaction is added to the tight-binding Hamiltonian to include local correlations, along with the Anisimov double counting correction \cite{Anisimov1991}.
For the Ni impurity, we use intra-orbital interaction $U=4.4$\,eV and Hund's coupling $J=0.65$\,eV; for Nd $U=2.5$\,eV and $J=0.25$\,eV. The interaction values were adopted from the constrained-random-phase approximation (cRPA) calculations in Ref.~\cite{Si2020}. Assuming spherical symmetry of the interaction, the inter-orbital coupling for both impurities is $U'=U-2J$. The interstitial-$s$ state is treated as uncorrelated (non-interacting).

%\lvc{Mott-vs-Hund's paper with Liang Si}. \ls{the paper is not published, shall we cite some other papers here?} \\
% TODO (Leonard): cite the Mott-vs-Hund's paper with Liang Si here once the bib entry exists.
Our calculations are performed at $\beta=400$\,eV$^{-1}$, corresponding to $T\approx29$\,K, which is close to the experimental temperature of 25\,K \cite{li2025}.
The impurity problem is self-consistently solved using continuous-time quantum Monte Carlo in hybridization expansion \cite{Gull2011a} on the Matsubara axis, as implemented in \textsc{W2DYNAMICS} \cite{w2dynamics2018}.\\
Local self-energies are analytically continued to real frequencies, using the maximum entropy method \cite{PhysRevB.57.10287,PhysRevB.44.6011}, as implemented in \textsc{ana\_cont} \cite{Kaufmann2021}. From that we calculate the DMFT lattice Green's function $\bm{G}(\bm{k}, \omega)$ using the Dyson equation.
In addition to the DMFT spectral function $A_{\alpha\beta} = -{1}/{\pi}\times (\mathrm{Im}\,\bm{G})_{\alpha\beta}$, we also consider photoemission matrix elements to model the ARPES signal directly in our localized Wannier basis, allowing for a direct comparison with experiment \cite{Matho2001ARPES}. This approach is often coined Wannier-ARPES \cite{PhysRevX.12.011019,g9d4-qls9,Yen2024}, and is extensively described in \cite{Yen2024}, though without DMFT correlations.
In multi-orbital systems, the measured intensity reflects a nontrivial interplay between the full spectral-function matrix and orbital-dependent dipole matrix elements \cite{dtg3-66t9,Yen2024,Matho1995Photocurrents}.
Using Fermi's golden rule and the sudden approximation, the multiband ARPES photocurrent is determined by 
\begin{align}
    I(\bm{k}_\parallel, E) \sim \sum_{\alpha,\beta} {M}^*_\alpha(\bm{k}_\parallel, E) {M}_\beta(\bm{k}_\parallel, E) {A}_{\alpha \beta}(\bm{k}, \omega) \label{eq:ARPES_signal_main}
\end{align}
where $M_\alpha$ is the dipole matrix element corresponding to the Wannier state $|\phi_{\bm{0}\alpha}\rangle$ in the atom-centered approximation (see Sec.~\ref{Sec:SM_arpes_intensities} in the SM \cite{SM} for details).
The crux of calculating $M_\alpha = \langle \chi_{\bm{k}_\parallel,E}|\bm{\varepsilon}\cdot\bm{r}
|\phi_{\bm{0}\alpha} \rangle$, where $\bm{\varepsilon}$  is the polarization vector of the incoming photons and $\bm{r}$ the position operator,
lies in the final photoelectron state $|\chi_{\bm{k}_\parallel,E} \rangle$---a high-energy state of the full system.
The conservation of in-plane momentum and energy sets a relationship between the photon energy $E$, the initial momentum $\bm{k} = (\bm{k}_\parallel, k_z)$, the photoelectron momentum and quasiparticle energy $\omega$. The photon energy $E$ dictates here the $k_z$ value probed in the photoemission experiment.\\
In this work, we employ the plane-wave final state approximation (PWA) \cite{Williams1977,Puschnig2009} $\chi_{\bm{p}} = \mathrm{e}^{i\bm{p}\cdot\bm{r}}$. 
This is a crude approximation in which the overall shape and symmetries of matrix elements are qualitatively well described; however, it might fail to quantify relative intensities between orbitals accurately.
Soft x-ray photon energies of ${\sim}100$--$1000$\,eV correspond to electron mean-free paths of up to several nanometers \cite{Seah1979}, providing a somewhat greater bulk sensitivity than ultraviolet (UV) ARPES.
The experimental ARPES images of NdNiO$_2$ \cite{li2025} were obtained using synchrotron radiation in the soft x-ray regime, with a photon energy of $E = 160$\,eV. Assuming a plane-wave final state, this corresponds to probing the Fermi surface at $k_z \sim 7\frac{\pi}{c}$ (see Sec.~\ref{Sec:SM_kz_determination} in the SM \cite{SM} for further discussion).\\

\section{Results}
\label{Sec:results}

\subsection{A-pocket: hybridizations}
\label{Sec:results_DMFT}

\begin{figure*} [tbp]
\centering
\includegraphics[width=\textwidth]{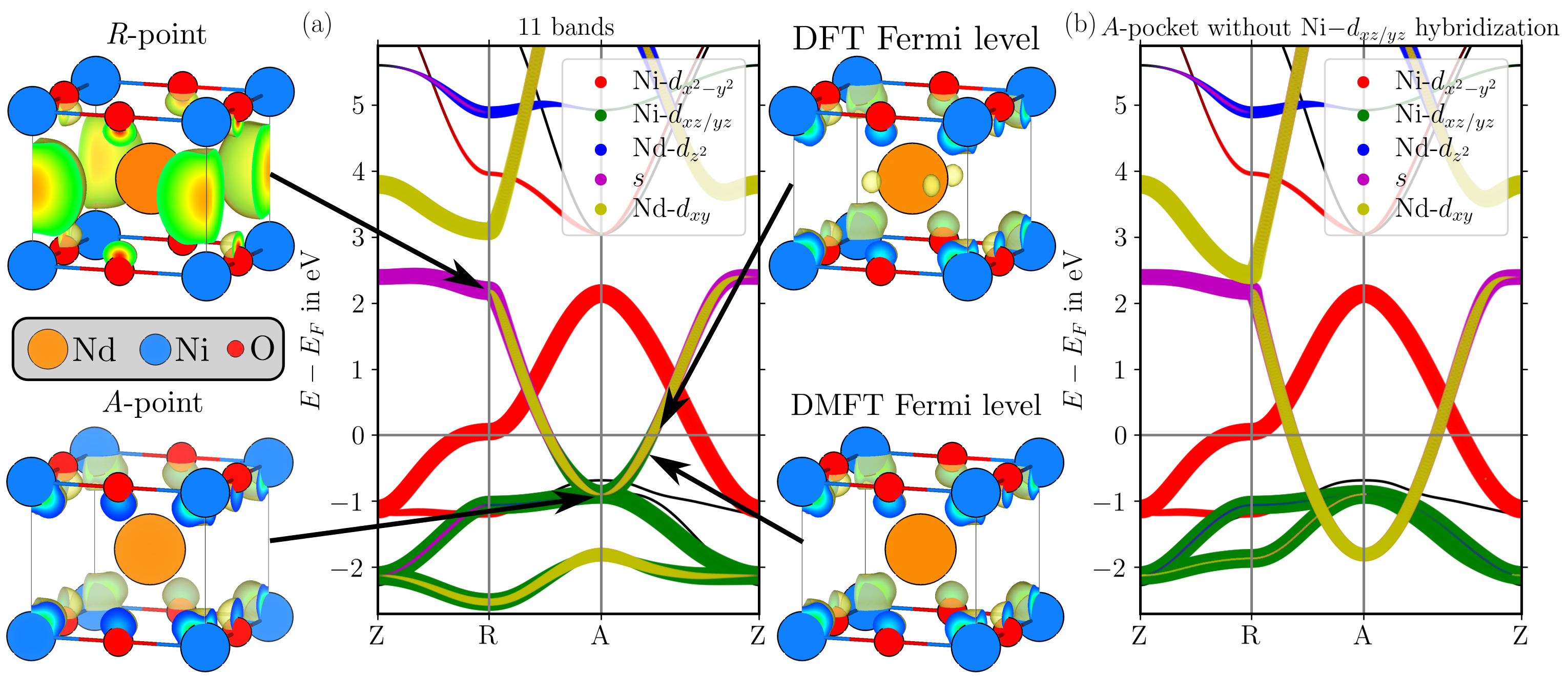}
\caption{\label{fig:DFT_bands} DFT bands and projections onto local orbitals for the 11-band Wannier model in the $k_z=\pi$ plane. (a) Full 11-band model. Side panels show DFT-calculated charge density for the $A$-pocket band at different $\bm{k}$-points indicated by arrows. (b) Hopping from Ni-$d_{xz/yz}$ to interstitial-$s$ and Nd-$d_{xy}$ is set to 0 such that the $A$-pocket does not hybridize with Ni on its way down.}
\end{figure*}

\begin{figure} [tbp]
        \centering
        \includegraphics[width=\columnwidth]{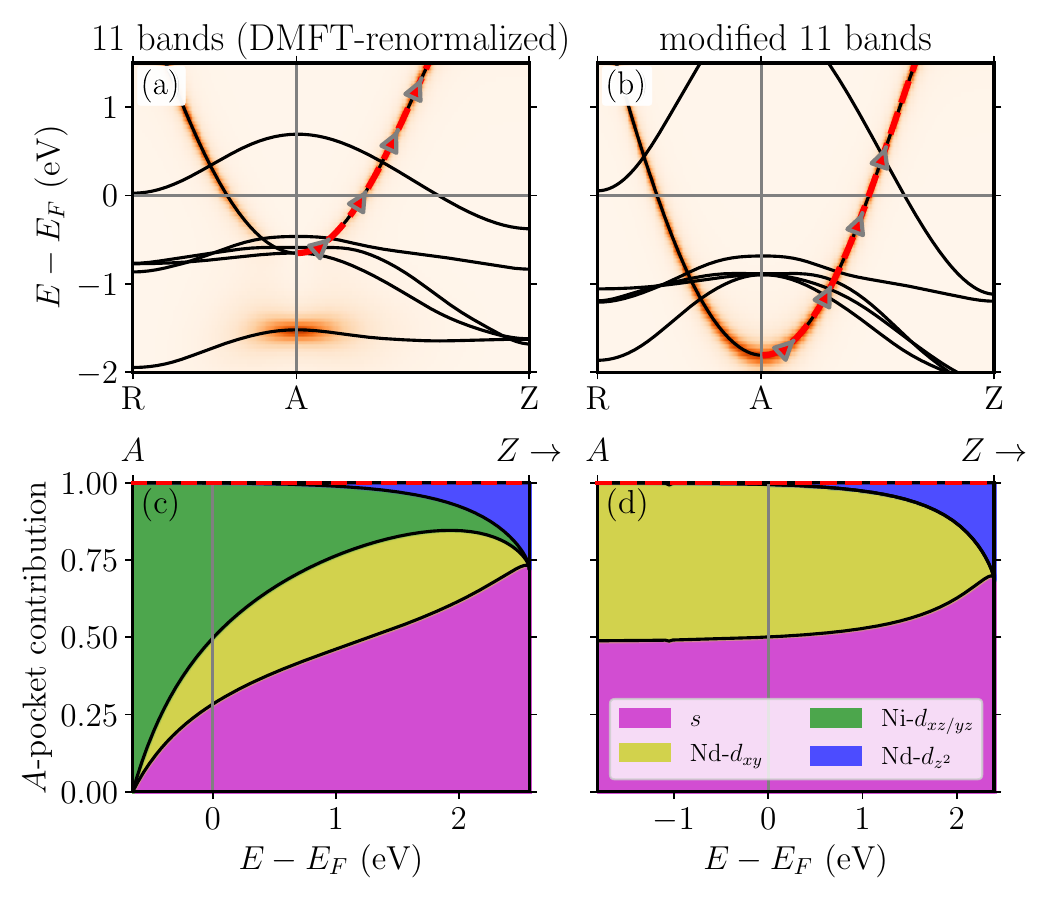}
        \caption{\label{fig:DFT_pros} Zoom-in on the $A$-pocket
        for (a) 11-band model and (b) artificial model without hybridization of  Nd-$d_{xy}$ and interstitial-$s$ 
        to Ni-$d_{xz/yz}$ orbital.
        The interstitial-$s$ orbital character is shown as the colormap.
        (c) and (d) Quantitative contribution of relevant atomic orbitals to the $A$-pocket 
        for the two models of (a) and (b) as a function of energy along the path indicated by gray arrows in (a,b). While the DFT and DMFT-renormalized bands contain Ni-$d_{xz/yz}$, Nd-$d_{xy}$, and interstitial-$s$ contributions around $E_F$, the pocket in the artificial model in (b,d) is formed by Nd-$d_{xy}$ and interstitial-$s$ states. In the 10-band model, which excludes the interstitial-$s$ state, the corresponding band contains Ni-$d_{xz/yz}$ and Nd-$d_{xy}$ contributions (see SM Sec.~\ref{Sec:SM_a_pocket_models}).}
\end{figure}

%The orbital origin of the electron-like pocket at the corner of the three-dimensional Brillouin zone, i.e., the $A$-point, remains under debate.
We address the orbital character of the $A$-pocket by first-principles DFT and DMFT calculations.
According to the DFT band structure shown in Fig.~\ref{fig:DFT_bands}, the $A$-pocket stems from a highly hybridized bonding state approximately 2\,eV above the Fermi level with a corresponding antibonding partner near 4\,eV. 
From the $Z$- to the $R$-point, this band is nearly flat, with charge localized in the interlayer region between two Ni sites, and the orbital character clearly being interstitial-$s$ [purple in Fig.~\ref{fig:DFT_bands}(a)].
However, when moving towards the $A$-point, hybridizations with Ni-$3d$ and Nd-$5d$ states come into play, causing the charge to become delocalized. At the same time this makes the corresponding band dispersive, even dipping well below the Fermi level.
From the charge density plots, shown as side panels to Fig.~\ref{fig:DFT_bands}(a) at different energies, it becomes evident that O-$p$ orbitals also play a role around the Fermi level. While oxygen-centered Wannier functions are not explicitly included in our low-energy Hamiltonian, the Ni and Nd $d$-like Wannier orbitals contain a certain portion of O-$p$ weight, which distinguishes them from pure atomic orbitals.

In Fig.~\ref{fig:DFT_pros}, the orbital character of the $A$-pocket at different $\bm{k}$-points is quantified by projecting the corresponding Bloch state onto the relevant Wannier orbitals. Exactly at the $A$-point the band contains only Ni-$d_{xz/yz}$ character. Following the band to higher energies, the Nd-$d_{xy}$ and interstitial-$s$ characters play an increasingly important role.
At the DFT Fermi level (see SM \cite{SM} Sec.~\ref{Sec:SM_a_pocket_models}), spectral weight of the $A$-pocket band is distributed among interstitial-$s$ (30\%), Nd-$d_{xy}$ (25\%) and Ni-$d_{xz/yz}$ (44\%) character. When considering the small upward shift of the $A$-pocket in DMFT,
these contributions change slightly to 30\%, 22\%, and 48\%, respectively, as shown in Fig.~\ref{fig:DFT_pros}(c).
Since the $A$-pocket is weakly correlated and merely shifts rigidly upward in DMFT spectra, we evaluate its orbital character from the DFT eigenstates at the DMFT Fermi crossing.
Also, because of the orbitals' symmetries, the $A$-pocket and Ni-$d_{x^2-y^2}$ band simply cross without any hybridization gap in the $k_z=\pi$ plane
(i.e., there is no hybridization between these orbitals when $k_z=\pi$).

To better understand the physics underlying the $A$-pocket, we construct an artificial Hamiltonian in which all interorbital hoppings between the Ni-$d_{xz/yz}$ orbitals and the interstitial-$s$ and Nd-$d_{xy}$ orbitals are set to zero. The resulting, decoupled bands are shown in Figs.~\ref{fig:DFT_bands}(b) and \ref{fig:DFT_pros}(b).
This simplified model reveals that the strong dispersion of the $A$-pocket band along the $R$--$A$ path arises from the hybridization between Nd-$d_{xy}$ and interstitial-$s$ orbitals. These orbitals form a bonding-antibonding pair, with the bonding state dispersing to about 2\,eV below the Fermi level.

In the full Hamiltonian [Fig.~\ref{fig:DFT_bands}(a)], the hybridization of both Nd-$d_{xy}$ and interstitial-$s$ with Ni-$d_{xz/yz}$ leads to a band reconstruction.
This raises the minimum of the $A$-pocket to  $-1$\,eV, and the orbital character at and around this minimum is mainly Ni-$d_{xz/yz}$.
The interstitial-$s$ and Nd-$d_{xy}$ character is instead pushed into the lower band of this reconstruction, whose top lies at about $-2$\,eV.
This reconstruction also explains why the $A$-pocket is much more robust than the $\Gamma$-pocket: Its minimum would be at a much lower energy of $-2$\,eV were it not for the reconstruction with the Ni-$d_{xz/yz}$ orbitals.
Shifting the minimum of the reconstructed $A$-pocket from $-1$\,eV to above the Fermi energy would hence require shifting the whole Ni-$d_{xz/yz}$ orbital manifold as well.
We have therefore identified the interstitial-$s$--Nd-$d_{xy}$ bonding state as the cause of the $A$-pocket.
However, hybridization with the Ni-$d_{xz/yz}$ band reshapes its minimum, its actual dispersion, and its orbital composition.

The interstitial-$s$ spectral contribution shown in the colormaps in Fig.~\ref{fig:DFT_pros}(a) and (b), i.e., with and without hybridization to  Ni-$d_{xz/yz}$, further supports this interpretation.
Without hybridization to the Ni-$d_{xz/yz}$ orbitals in Fig.~\ref{fig:DFT_pros}(d), the $A$-pocket is composed of approximately equal interstitial-$s$ and Nd-$d_{xy}$ contributions all the way from its minimum at $-2$\,eV to the $R$-point at about $+2$\,eV, reflecting its bonding-antibonding character.
However, when the hybridization to the Ni-$d_{xz/yz}$ is properly included in Fig.~\ref{fig:DFT_pros}(c), the orbital character changes. It is 100\% Ni-$d_{xz/yz}$ at the pocket minimum ($-1$\,eV), and Ni-$d_{xz/yz}$ remains the largest contribution up to the Fermi level. Above the Fermi level, i.e., when moving further away from the genuine Ni-$d_{xz/yz}$ orbital [originally between $-2$\,eV and $-1$\,eV in Fig.~\ref{fig:DFT_bands}(b)] towards the interstitial-$s$ band located at about $+2$\,eV without hybridizations, the character changes increasingly to interstitial-$s$.\\

To summarize the DFT(+DMFT) picture, interstitial-$s$ states do play a major role in the $A$-pocket, but Nd-$d_{xy}$, Ni-$d_{xz}$, and Ni-$d_{yz}$ orbitals also contribute significantly, especially at and below the Fermi energy. 
The strongly hybridized $A$-pocket is in apparent contradiction to the conclusion drawn in the first ARPES experiments on superconducting NdNiO$_2$ \cite{li2025} that the $A$-pocket is purely or dominantly interstitial-$s$. To resolve this controversy, we further simulate the ARPES signal directly and examine how the orbital character manifests---or is hidden---in the measured photocurrent.\\

\subsection{Simulated ARPES signal}
\begin{figure} [tbp]
        \centering
        \includegraphics[width=\linewidth]{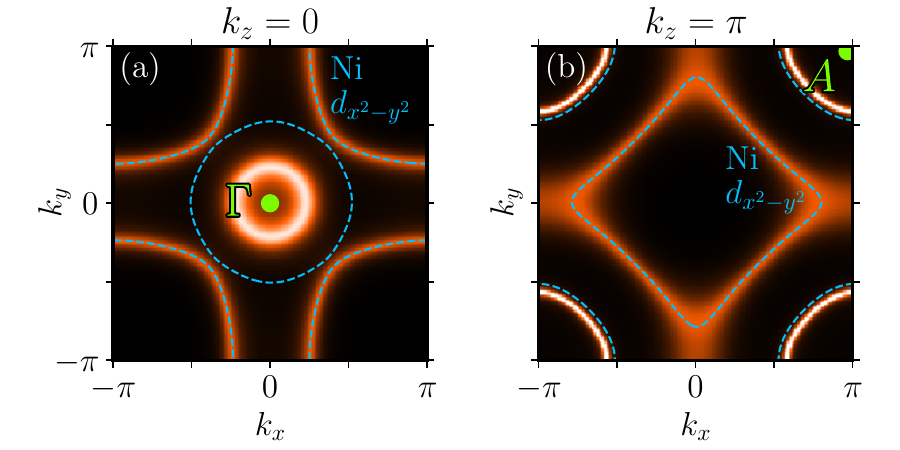}
        \caption{\label{fig:DMFT_FS} DFT Fermi surface (blue-dashed lines) for (a) $k_z=0$ and (b) $k_z=\pi$. Colormap shows DMFT Fermi surface with correlations renormalizing and smearing the bands. In both cases we have a Ni-$d_{x^2-y^2}$ Fermi surface, and a pocket around the $\Gamma$- and $A$-momentum.}
\end{figure}
In order to follow the reasoning of Li \emph{et al.} \cite{li2025}, we simulate the ARPES signal according to Eq.~(\ref{eq:ARPES_signal_main}), which lets us track how individual Fermi surface features appear or disappear for different polarizations and geometries.
Fig.~\ref{fig:DMFT_FS} shows the full bulk Fermi surface, calculated in DMFT as $\mathrm{tr}\{\bm{A}(\bm{k}, \omega=0)\} = \sum_\alpha A_{\alpha\alpha}(\bm{k}, \omega=0)$ at (a) $k_z=0$ and (b) $k_z=\pi$. The DFT Fermi contour is superimposed as blue dashed lines. Note that our DFT calculations find an electron-like pocket in the $k_z=0$-plane around the BZ center $\Gamma$. Although the corresponding band is strongly shifted upward in DMFT, a small pocket is still present, whereas none is observed experimentally \cite{li2025}. In contrast, for Sr-doped NdNiO$_2$ and LaNiO$_2$ it is absent in both DMFT calculations \cite{Kitatani2020} and experimental measurements \cite{Ding2024,Sun2025}.
Furthermore, the local correlations included in DMFT cause a renormalization and visible smearing of the almost half-filled Ni-$d_{x^2-y^2}$ orbital.
In DFT, the Ni-$d_{x^2-y^2}$ orbital forms an electron-like pocket around the $Z$-point, whereas it appears hole-like in DMFT due to both broadening and correlation shifts.
In addition, the $A$-pocket is shifted to slightly higher energies, resulting in a smaller pocket compared to DFT.\\

\begin{figure} [tbp]
        \centering
        \includegraphics[width=\columnwidth]{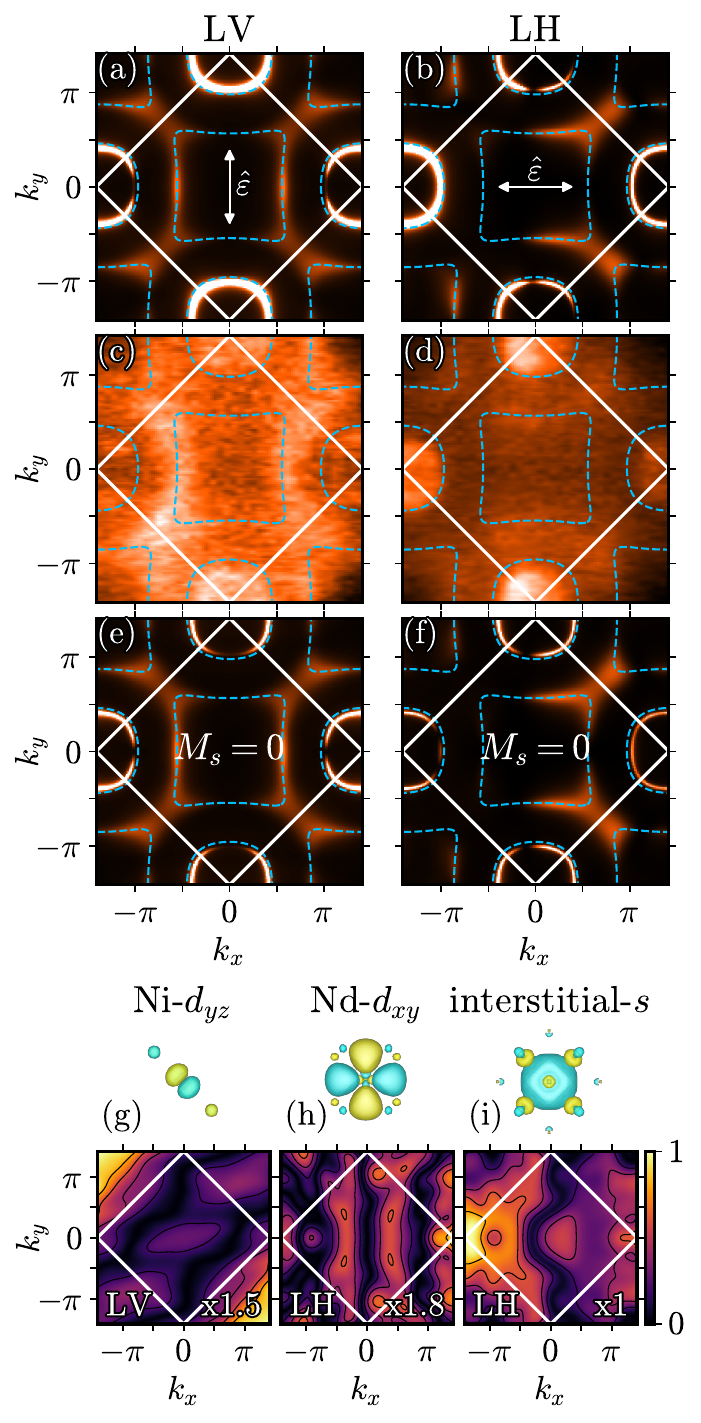}
        \caption{\label{fig:ARPES_rot}Fermi surface maps in the rotated basis at $k_z=7\pi/c$. Left: linear vertical, right: linear horizontal polarization. (a) and (b): our simulated ARPES images in the 11-band model. (c) and (d): Experimental images recreated from \cite{li2025}. (e) and (f): Simulated ARPES images in the 11-band model, with $M_s \equiv 0$ artificially enforced. (g)--(i): Matrix elements, including a top-view of the corresponding Wannier function above the panels. The incident angle is 45°, relevant for the normal component of polarization in LH. White squares indicate the first Brillouin zone, and blue dashed lines are the DFT Fermi contour. Note that in the rotated basis, following the notation of \cite{li2025}, $k_x$ and $k_y$ are no longer along the cubic axes.}
        
\end{figure}

\begin{figure} [tbp]
        \centering
        \includegraphics[width=\columnwidth]{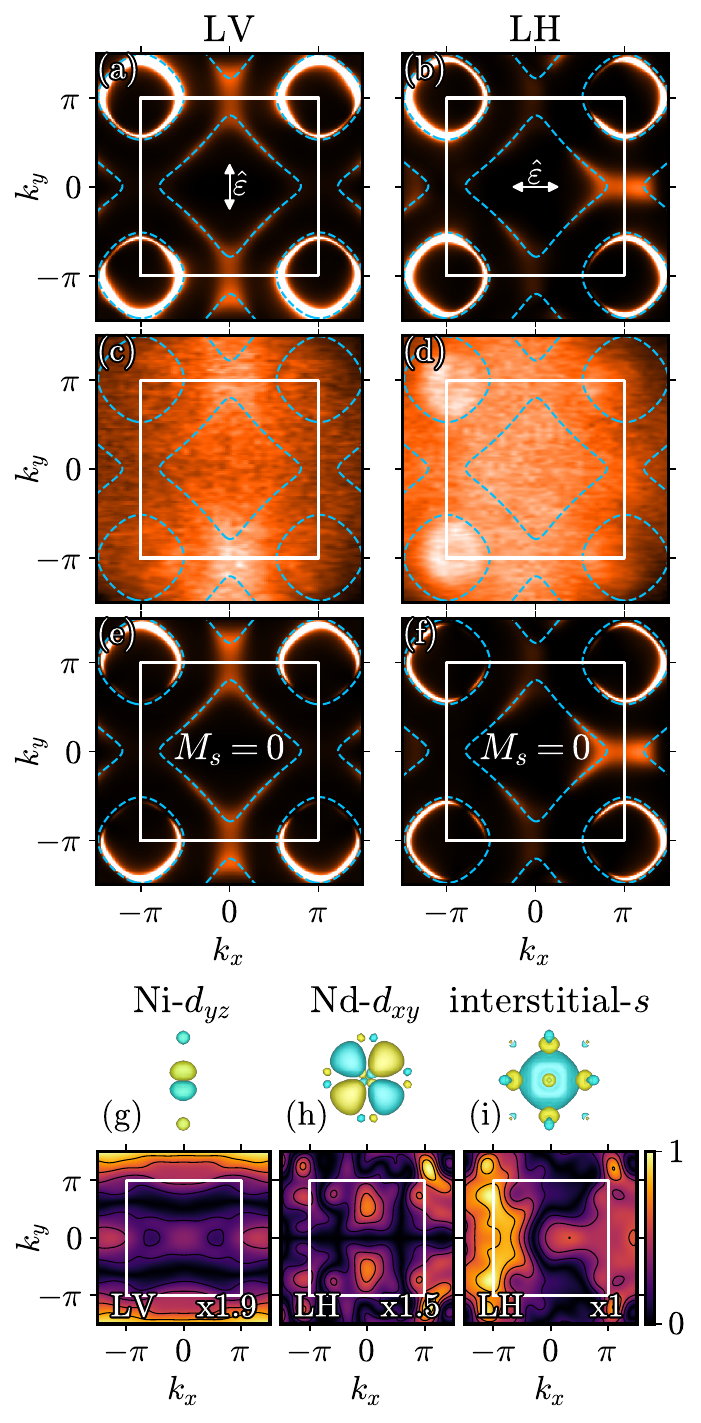}
        \caption{\label{fig:ARPES}Same as Fig.~\ref{fig:ARPES_rot}, but measured in the conventional basis in which the Ni--O bonds point along the $x$- and $y$-directions.}
\end{figure}

Due to the four-fold symmetry ($D_{4h}$) of NdNiO$_2$, the Fermi surface shows the same rotational symmetry within the Brillouin zone. Linearly polarized light breaks this symmetry, so that the measured maps have a lower symmetry than the Fermi surface and crystal structure.
These effects are covered in Figs.~\ref{fig:ARPES_rot} and \ref{fig:ARPES}, which show the Fermi surface at $k_z=7\pi/c$ (see SM  \cite{SM} Sec.~\ref{Sec:SM_kz_determination}) calculated according to Eq.~(\ref{eq:ARPES_signal_main}) in different experimental geometries for linear horizontally (LH; $\hat\varepsilon$ parallel to incidence plane) and linear vertically (LV; $\hat\varepsilon$ perpendicular to incidence plane) polarized photons (see SM Sec.~\ref{Sec:SM_experimental_setup} for the experimental setup).\\

%discussion of rotated frame

The experimental study \cite{li2025} first presents measurements with the sample rotated by 45° around its normal axis (relative to the standard basis, where Ni--O bonds define the Cartesian $x$- and $y$-direction).
Our simulations for this rotated setup are presented in Fig.~\ref{fig:ARPES_rot}:
As expected, the Fermi surface maps calculated for (a) LV and (b) LH look different and have reduced symmetries that match the measured Fermi surface maps  in panels (c) and (d), respectively (recreated from the data provided through the repository \cite{Li2025Data} from \cite{li2025}).
We find the theoretical intensities of the Fermi surface features to be in qualitative agreement with experiment: This includes, first, the dependence of the Ni-$d_{x^2-y^2}$ intensity on the polarization direction, and second the enhancement and suppression of the $A$-pocket at the corner of the Brillouin zone (given by the white, rotated square).

One strength of the three-step model is that the photocurrent can be decomposed into orbital channels. In order to see the importance of interstitial-$s$ states for the images, we thus further calculate the photocurrent by enforcing $M_s \equiv 0$ in Fig.~\ref{fig:ARPES_rot} (e) and (f), i.e., suppressing the $M_s$ channel. This removes the interstitial-$s$ contribution. For these images, the $A$-pocket intensities no longer match the experiment. In particular, the spectral weight of the $A$-pocket in Fig.~\ref{fig:ARPES_rot} (e) gets strongly suppressed for $k_x=0$.
In LH, both Fig.~\ref{fig:ARPES_rot} (b) and (d) show asymmetric intensities of the $A$-pocket, with a brighter signal for $k_x < 0$ than for $k_x > 0$. When neglecting $M_s$ in panel (f), this effect is lifted. Hence, the asymmetry stems from the interstitial-$s$ orbital contribution, as we can also see from the matrix element plotted in panel (i).
We take the sensitivity of $A$-pocket intensity to $M_s$ as evidence that the interstitial-$s$ state is essential to the measured photocurrent.\\

Suppressing the Nd-$d_{xy}$ or Ni-$d_{xz/yz}$ channels instead, i.e., enforcing $M_{d_{xy}}\equiv 0$ or $M_{d_{xz/yz}}\equiv 0$, leaves the $A$-pocket intensity almost unchanged [SM Sec.~\ref{Sec:SM_channels_dark}, Fig.~\ref{fig:SM_no_d}]. The reason is not a small orbital weight but a small matrix element: at the corner of the Brillouin zone $M_{d_{xy}}$ and $M_{d_{xz/yz}}$ are much smaller than $M_s$ [see Figs.~\ref{fig:ARPES_rot}(g)--(i)]. Therefore, the intensity of the $A$-pocket stems mostly from the interstitial-$s$ channel, although this orbital provides only about 30\% of the spectral weight of the state. In other words, the interstitial-$s$ orbital dominates the \emph{photoemission signal}, not the \emph{wave function}. Based purely on non-hybridized, atomic-like $d$ orbitals, as considered in SM Sec.~\ref{Sec:SM_atomic_matrix_elements}, the contribution of the $d$ orbitals would be large.
%this effect cannot be observed. 
This immediately explains why Li et al.~\cite{li2025} observed a pocket that behaves like a pure $s$ state. ARPES in this geometry is essentially blind to the Ni-$d_{xz/yz}$ and Nd-$d_{xy}$ admixture and the measured maps are fully consistent with the hybridized pocket of Sec.~\ref{Sec:results_DMFT}, but they can neither confirm nor exclude it.

%Here comes discussion of non-rotated frame

Simulations and measurements in the non-rotated basis are presented in Fig.~\ref{fig:ARPES}. Keeping the same polarization directions as before in the laboratory frame, a rotation of the sample effectively alters the polarization, providing additional insights about the constituents that make up the Fermi surface.
Again, the symmetries of simulated Fermi surface maps match well to the measured ones. However, let us note a mismatch between calculation and measurement: Especially in LV, the experiment [Fig.~\ref{fig:ARPES} (c)] obtains a much higher intensity stemming from the Ni-$d_{x^2-y^2}$ band, compared to the $A$-pocket. This is not surprising: While the PWA is able to correctly cover orbital symmetries, it is not reliable for relative intensities between different orbitals at different momenta. The intensity distribution within the $A$-pocket discussed above is, however, governed by the symmetry of the matrix elements in a narrow momentum region and is therefore less sensitive to these details.\\

As the interstitial-$s$ orbital's in-plane symmetry axes change only slightly upon the 45° rotation, $M_s$ has a similar asymmetry in both the rotated and non-rotated frame, i.e., a larger weight for $k_x < 0$ than for $k_x>0$. Similar to the calculations in the rotated frame (Fig.~\ref{fig:ARPES_rot}), when turning off the interstitial-$s$ channel in the photocurrent, the distribution of $A$-pocket intensities in Fig.~\ref{fig:ARPES} (e,f) no longer matches the experiment, which becomes most prominent in LH polarization.

%Now, follow their reasoning why strong contribution should come from interstitial-$s$} 

Based on their measurements \cite{li2025}, Li \emph{et al.} systematically exclude the possible contributions of Ni- and Nd-$d$ orbitals proposed in \cite{Si2024,Gu2020,Nomura2019,Wu2020,Adhikary2020,Worm2024} and also established in this work (see Sec.~\ref{Sec:results_DMFT}) from the Fermi surface. Their analysis treats the ARPES signal as a weighted trace of the intrinsic spectral function (Fig.~\ref{fig:DMFT_FS}), with each orbital weighted by the squared matrix element of the corresponding atomic-like $3d$ orbital from Ref.~\cite{Ye_2013}.
Our matrix elements for atomic-like $3d$ orbitals are depicted in Fig.~\ref{fig:ME_examples} and qualitatively agree with \cite{Ye_2013}.
Note that \cite{Ye_2013} does not consider an interstitial-$s$ state, which Li et al. approximate by the $3d_{z^2}$ orbital due to their similar in-plane symmetry \cite{li2025}. However, the two rightmost columns of Fig.~\ref{fig:ME_examples} show that the interstitial-$s$ and $d_{z^2}$ orbitals respond differently to the out-of-plane polarization component in LH.\\

% Why Nd-dxy cannot play a role
In more detail, the authors of \cite{li2025} first exclude a contribution from Nd-$d_{xy}$, based on the magnitude of its matrix element. Along a vertical $A$--$Z$--$A$ cut, for the in-plane component $\hat{\varepsilon}=(1,0,0)$ of LH polarization, the matrix element exactly vanishes due to symmetry for both the perfect atomic-like orbital in Ref.~\cite{Ye_2013} and Supplemental Fig.~\ref{fig:ME_examples}, as well as for the hybridized Wannier orbital. For LH polarization, with an out-of-plane component, this condition is weakened, modifying the path of $M_{d_{xy}}=0$ slightly [Fig.~\ref{fig:ARPES_rot} (h)]. The symmetry of the (atomic and Wannier) orbital implies that there should be a strong suppression of the Nd-$d_{xy}$ channel in the ARPES image measured along or close to this cut. The fact that the pocket is still visible in Fig.~\ref{fig:ARPES_rot} (b,d) disfavors Nd-$d_{xy}$ states alone as the origin of the $A$-pocket. However, this is not a contradiction to our finding in Sec.~\ref{Sec:results_DMFT}, that Nd-$d_{xy}$ provides about 22\% of spectral weight to the $A$-pocket at the Fermi level.
Next, the $A$-pocket in LV is much less intense than the hole-like Ni-$d_{x^2-y^2}$ pocket around the $Z$-point [Fig.~\ref{fig:ARPES}(c)].
This contradicts the large magnitude of the atomic-like $3d_{yz}$ matrix element reported in Ref.~\cite{Ye_2013} and shown in Supplemental Fig.~\ref{fig:ME_examples}.
This image is indeed the strongest deviation from our results, as the bright $A$-pocket is actually seen in Fig.~\ref{fig:ARPES}(a)---since the matrix element of the interstitial-$s$ orbital in Fig.~\ref{fig:ARPES} (i) is large at the Brillouin zone corners.

Further, the Ni-$d_{xz/yz}$ Wannier orbital is hybridized with O-$p_z$, strongly modulating its matrix element in Fig.~\ref{fig:ARPES}, compared to the atomic-like matrix element. Together with the small magnitude of $M_{d_{xz/yz}}$ near the $A$-pocket noted above, this undermines an exclusion argument based on the large atomic-like matrix element.
However, due to absolute intensity effects not covered in the plane-wave final-state approximation, we cannot further examine if the discrepancy in relative intensities is a flaw of the spectral composition in DFT+DMFT or an effect masked by matrix elements that are beyond the plane-wave final-state approximation.

The simplified, weighted trace description is a good starting point. However, it has two flaws: First, the reduction to the weighted trace is generally not valid in multi-orbital systems, because (complex, possibly negative) off-diagonal elements of the spectral function combine coherently with orbital-dependent matrix elements and can produce interferences \cite{dtg3-66t9,Yen2024,Beaulieu2020,Matho1995Photocurrents}.
Second, as we can see from the bottom rows in Figs.~\ref{fig:ARPES_rot} and \ref{fig:ARPES}, the relevant Wannier orbitals forming the basis for a low-energy description of NdNiO$_2$ strongly deviate from perfect atomic-like orbitals, mostly due to inter-orbital hybridizations with O-$2p$. This can completely alter the matrix elements and hence the interpretation of the photocurrent.\\

Overall, the magnitudes of matrix elements for the interstitial-$s$ orbital throughout the Brillouin zone are consistent with the measured and simulated intensities of the $A$-pockets. This, together with setting $M_s=0$, allows us to conclude that the interstitial-$s$ orbital provides much of the $A$-pocket's ARPES signal, but the sizable Nd-$d_{xy}$ and Ni-$d_{xz/yz}$ contributions found in DFT+DMFT are compatible with the ARPES maps at this geometry. The ARPES images in Ref.~\cite{li2025} therefore do not contradict the hybridized $A$-pocket found in DFT and DFT+DMFT.

\section{Conclusion}
\label{Sec:conclusion}
Our DFT+DMFT calculations identify the electron pocket state of infinite-layer nickelates around the $A$-momentum as a strongly hybridized state with substantial interstitial-$s$, Nd-$d_{xy}$, and Ni-$d_{xz/yz}$ character near the Fermi energy.
This $A$-pocket band crosses the Fermi energy in the first place because the bonding combination of the interstitial-$s$ and Nd-$d_{xy}$ orbitals leads to a strong dispersion.
Complicating things, this bonding combination further hybridizes and reconstructs with the Ni-$d_{xz/yz}$ band. Because of all of this, the $A$-pocket is (i) at its bottom---at the $A$-momentum---predominantly of Ni-$d_{xz/yz}$ character; (ii) around the Fermi energy a strong admixture of Ni-$d_{xz/yz}$, interstitial-$s$, and Nd-$d_{xy}$ character; and (iii) towards $+2$\,eV (the point of origin of the interstitial-$s$ orbital) predominantly interstitial-$s$ character. Local DMFT correlations strongly renormalize the Ni-$d_{x^2-y^2}$ band, but shift the $A$-pocket only slightly upward in energy, otherwise hardly changing its dispersion. Correlations thus only weakly modify the orbital composition of the $A$-pocket at the Fermi level.
    
Our DFT+DMFT simulations of the ARPES spectra well reproduce the principal polarization-dependent symmetries and the qualitative intensity variations of the $A$-pocket in the measured Fermi surface maps \cite{li2025}.
In agreement with \cite{li2025}, we find that within our approximation
the interstitial-$s$ component is essential to reproduce the polarization-dependent ARPES features of the $A$-pocket.
However, neither the experimental nor the simulated data require the $A$-pocket to have purely or predominantly interstitial-$s$ character. Instead, the apparent pure-$s$ character of the pocket in ARPES is due to a matrix-element effect. Under the experimental conditions in Ref.~\cite{li2025}, the Ni-$d_{xz/yz}$ and Nd-$d_{xy}$ contributions are suppressed compared to the interstitial-$s$ state. 
Hence, the interstitial-$s$ orbital contributes much more to the ARPES signal at the $A$-pocket than its admixture to the wavefunction would suggest.
%the ARPES signal is dominated by interstitial-$s$ while the state itself is not. 
Artificially suppressing the $d$-derived matrix elements in our simulation accordingly leaves the Fermi surface maps almost unchanged.
We emphasize that the matrix elements must be evaluated with the actual Wannier functions rather than with atomic orbitals.
% For resolving the orbital composition of the $A$-pocket experimentally would therefore require photon energies or polarization geometries in which the $d$-derived matrix elements are not suppressed, which the present approach can help to identify.
% \lvc{propose other geometries/photon energies where we can identify the differences?}
% \khc{We would need to. Otherwise the last sentence is maybe better omitted .}
% \khc{Why nothing on p contribution in conclusion?}

\section{Acknowledgments}
We thank Viktor Christiansson, Eric Jacob, Andriy Smolyanyuk, J\'an Min\'ar, and Konrad Matho for helpful discussions.
We further acknowledge funding by the Austrian Science Fund (FWF) through project DOI 10.55776/I5398 and the European Research Council (ERC) through ERC-2024-ADG RealSuper project DOI 10.3030/101201037. The DFT+DMFT calculations were performed primarily on the Vienna Scientific Cluster (VSC).
L.~S.~acknowledges support from the National Natural Science Foundation of China (Grant Nos.~12422407 and 12674309) and the Key Research and Development Program of Shaanxi (2024QY2-GJHX-42).

This project is funded in part by the European Union. Views and opinions expressed are however those of the author(s) only and do not necessarily reflect those of the European Union or the European Research Council Executive Agency. Neither the European Union nor the granting authority can be held responsible for them.

For the purpose of open access, the authors have applied a CC BY public copyright license to any Author Accepted Manuscript version arising from this submission.

\section{Data availability}
The input and output data for the calculations presented in the main text will be made openly available in a public repository upon publication of this article. The code for calculating photoemission matrix elements using either maximally localized Wannier functions obtained with Wannier90 \cite{Pizzi2020} or atomic-like orbitals will also be released in a public repository under the GPL-3.0 License.

\bibliography{main}
\clearpage

%%%% XXXXXXXXXXXXXX %
\pagebreak
\widetext
\newpage
\begin{center}
    \textbf{
    \huge
    Supplementary material for ``Orbital character and photoemission signatures of the $A$-pocket in infinite-layer nickelates''}
\end{center}
\vspace{1cm}
This supplemental material provides additional methodological details and results. Section~\ref{Sec:SM_wannierization} presents the Wannierization and compares the 10- and 11- band models with the DFT band structure. Section~\ref{Sec:SM_a_pocket_models} examines the $A$-pocket and its orbital character in different models. Section~\ref{Sec:SM_arpes_intensities} derives the expressions used to calculate ARPES intensities, and Section~\ref{Sec:SM_kz_determination} describes the determination of the probed $k_z$. Section~\ref{Sec:SM_matrix_elements} compares photoemission matrix elements for Wannier and atomic-like orbitals, validates our implementation against the independent \textsc{chinook} package, and discusses hybridization effects. Section~\ref{Sec:SM_channels_dark} complements the main text by switching off the $d$-derived matrix-element channels instead of the interstitial-$s$ channel. Section~\ref{Sec:SM_experimental_setup} describes the experimental geometry and light polarizations used in the simulations. Finally, Section~\ref{Sec:SM_lanio2_comparison} compares the band structure and orbital character of the $A$-pocket in LaNiO$_2$ with those in NdNiO$_2$ and shows the corresponding simulated Fermi surface maps.

\renewcommand{\thefigure}{S\arabic{figure}}
\renewcommand{\thesection}{S\arabic{section}}
\renewcommand{\thetable}{S\Roman{table}}
\renewcommand{\theequation}{S\arabic{equation}}
\renewcommand{\theHfigure}{S\arabic{figure}}
\renewcommand{\theHsection}{S\arabic{section}}
\renewcommand{\theHequation}{S\arabic{equation}}

\setcounter{equation}{0}
\setcounter{figure}{0}
\setcounter{table}{0}
\setcounter{section}{0}
\makeatletter
% make S1... before figures and equations
\renewcommand{\figurename}{Fig.}
\renewcommand{\thefigure}{S\arabic{figure}}
\renewcommand{\theequation}{S\arabic{equation}}
\renewcommand{\tablename}{Tab.}
\renewcommand{\thetable}{S\arabic{table}}

\newpage

\section{Wannierization}
\label{Sec:SM_wannierization}
\begin{figure}[tb]
    \centering
    \includegraphics[width=0.8\textwidth]{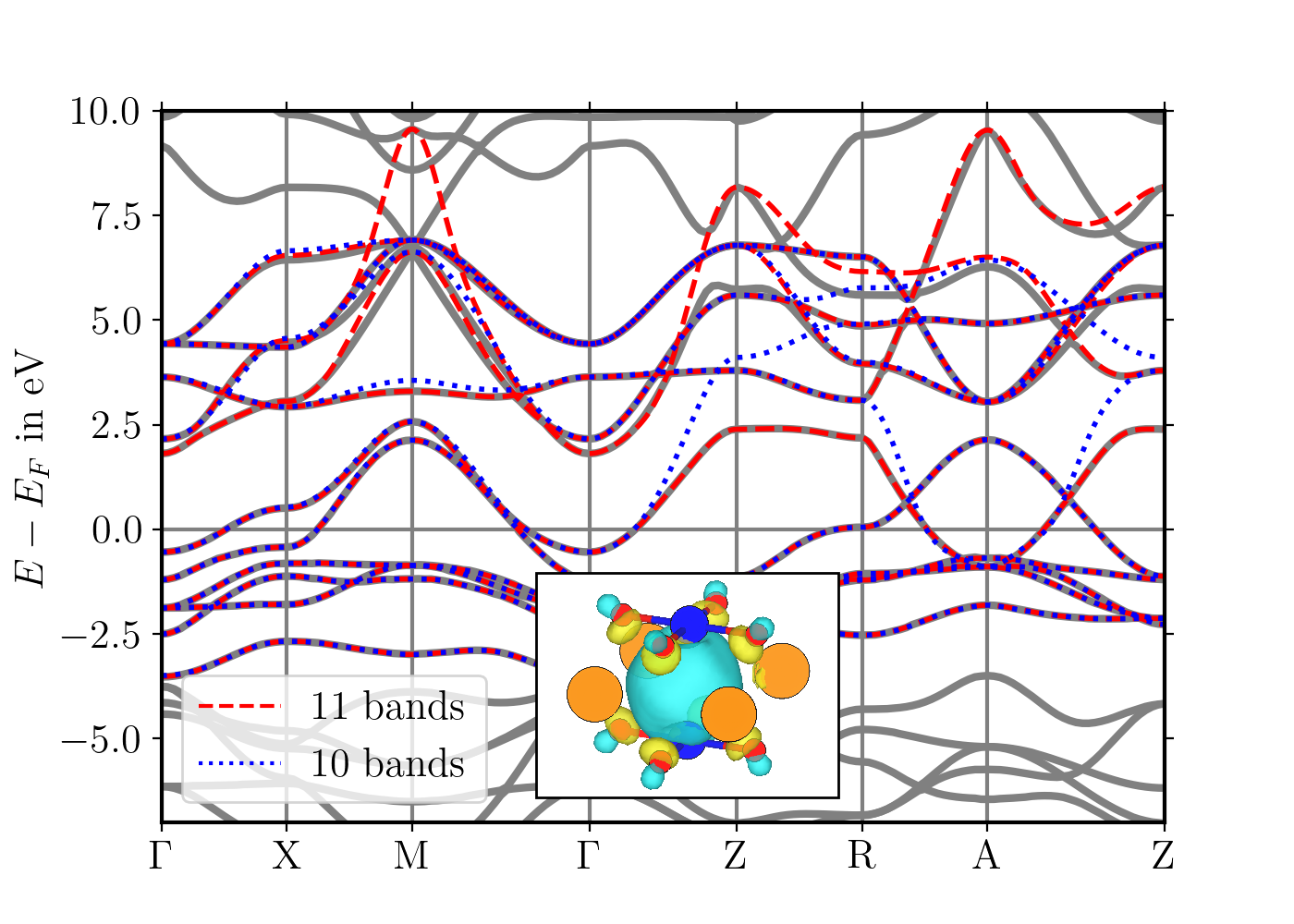}
    \caption{\label{fig:wannier_SM} DFT band structure of NdNiO$_2$ (gray lines) along a path through the whole BZ. Superimposed are Wannier interpolated bands for 11-band model (red; 5 Ni-$3d$, 5 Nd-$5d$, and interstitial-$s$) and 10-band model (blue; 5 Ni-$d$ and 5 Nd-$d$). For Wannierization of 11-band model we chose a frozen energy window of [-3.3, 0.5]\,eV. The inset plots the Wannier function corresponding to interstitial-$s$ orbital.}
\end{figure}

Fig.~\ref{fig:wannier_SM} plots the 
Wannier-projected bands on top of the full DFT
band structure. We show results for both the 10-band model (5 Ni-$d$ and 5 Nd-$d$) and the 11-band model (10-band + interstitial-$s$). The states close to Fermi level are well reproduced in both models. However, the 10-band model misses one orbital which necessarily leads to deviations at higher energies. This is, in particular, noticeable outside the $k_z=0$-plane above 1\,eV. The 11-band model does not have this weakness and successfully covers the $A$-pocket and DFT band structure up to 5\,eV. Note that the $A$-pocket in the 10-band model is not a smooth parabola [also visible in Fig.~\ref{fig:band_pros_SM} (d)], as in the 11-band model, which is also observable in the 10-band model used in \cite{Kitatani2020}. We observed that the exact shape of the $A$-pocket band---and up to which energy it reproduces the Bloch bands---strongly depends on the choice of frozen window.\\
The Wannier function corresponding to the interstitial-$s$ orbital is shown as an inset of Fig.~\ref{fig:wannier_SM}. It is localized in the interstitial region between 2 Ni atoms, where apical oxygen sits in the perovskite phase. As we can see, oxygen hybridizations also play a role, making it not perfectly $s$-shaped.
This deviation from atomic-like orbitals is also evident for the $d$-derived Wannier functions, since they substantially hybridize with O-$p$ orbitals, while the latter are not explicitly included in both of the presented Wannier models.

\section{\texorpdfstring{$A$}{A}-pocket in different models}
\label{Sec:SM_a_pocket_models}
\begin{figure}[tb]
    \centering
    \includegraphics[width=\textwidth]{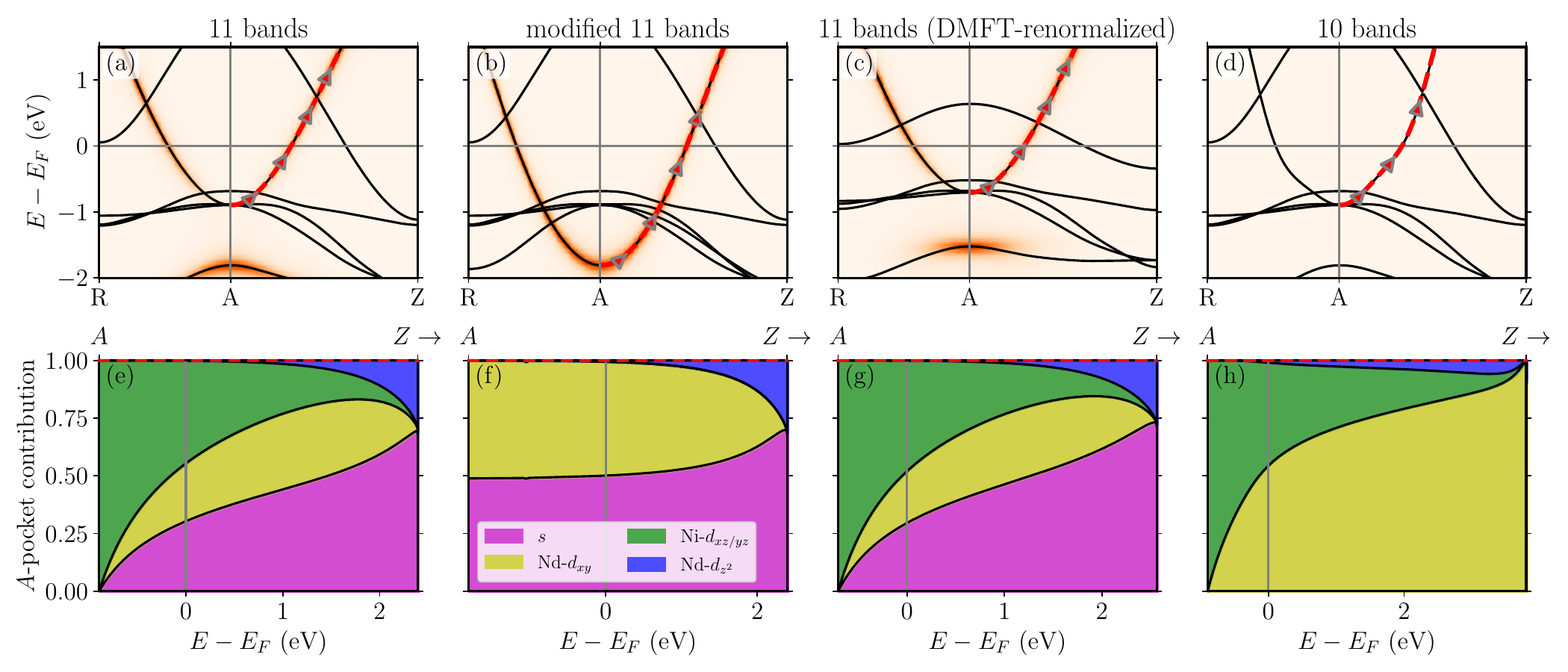}
    \caption{\label{fig:band_pros_SM} (a)--(d) Bands around $A$-pocket in different models. Color map in the background quantifies spectral weight of interstitial-$s$ orbital. (e)--(h) Corresponding atomic orbital contributions to $A$-pocket, going from band minimum to higher energies.}
\end{figure}
In Fig.~\ref{fig:band_pros_SM}, we compare the $A$-pocket in different tight-binding models, corresponding to the band structure of NdNiO$_2$. The DMFT-renormalized $A$-pocket slightly shifts upwards compared to DFT; however, the overall orbital contribution at Fermi level is similar.
In the modified 11-band model (see main text), hybridizations of $A$-pocket with Ni-$d_{xz/yz}$ are restricted and, consequently, the pocket is simply the bonding combination of interstitial-$s$ and Nd-$d_{xy}$ states.
The 10-band model does not consider interstitial-$s$ states. Hence, in Fig.~\ref{fig:band_pros_SM} (h), the missing weight is mainly provided by Nd-$d_{xy}$-derived states.

\section{Calculation of ARPES intensities}
\label{Sec:SM_arpes_intensities}
The framework we employ in the main text to simulate the ARPES signal is commonly termed Wannier-ARPES \cite{g9d4-qls9,Yen2024,PhysRevX.12.011019,Day2019}, where we mostly follow the conventions established in Ref.~\cite{Yen2024}. We assume a 2D system, e.g., a surface or slab consisting of many layers, and employ the atom centered approximation (ACA). All quantities are formulated in the basis of maximally localized Wannier functions $\phi_{\alpha\bm{R}}(\bm{r})$ \cite{RevModPhys.84.1419}, centered at $\bm{R}+\bm{r}_\alpha$ and normalized over all space, $\langle \phi_{\beta\bm{R}'}|\phi_{\alpha\bm{R}} \rangle = \delta_{\alpha\beta}\delta_{\bm{RR}'}$. This is the basis in which both the tight-binding Hamiltonian $\bm{H}^{\mathrm{DFT}}(\bm{k})$ obtained from Wannier90 and the DMFT self-energy are defined, and it is therefore the natural basis for the photoemission calculation.

Starting from Fermi's golden rule and applying the sudden approximation to decouple the photoelectron from the remaining system, the photocurrent is governed by dipole matrix elements between the occupied initial states and the photoelectron final state. We explicitly choose dipole gauge, $\hat{\Delta}=\bm{\varepsilon}\cdot\bm{r}$ with $\bm{\varepsilon}$ the polarization vector of the incoming photons, and a plane-wave final state $\langle \chi_{\bm{p}}|$ with three-dimensional momentum $\bm{p}$. Within the ACA \cite{Yen2024}, the photoemission amplitude of each Wannier orbital is evaluated as if the orbital were isolated at its center; formally, this amounts to neglecting terms proportional to $\nabla_{\bm{k}}(U^{\dagger(\bm{k})})_{\alpha n}$ (with $\bm{U}^{(\bm{k})}$ defined below). The elementary quantity entering the calculation is then the Wannier matrix element
\begin{align}
    M_\alpha(\bm{k},E) &= \langle \chi_{\bm{p}} | \hat{\Delta} | \phi_{\alpha} \rangle \nonumber\\
    &= \mathrm{e}^{-i\bm{p\cdot r}_\alpha}\, \mathrm{e}^{z_\alpha/\lambda}\int_{\mathbb{R}^3} d\bm{r}\, \mathrm{e}^{-i\bm{p\cdot r}}\, \bm{\varepsilon} \cdot \bm{r}\, \phi_{\alpha}(\bm{r+r_\alpha}). \label{eq:new_M_wannier}
\end{align}
Here, $\phi_\alpha(\bm{r})\equiv\phi_{\alpha\bm{0}}(\bm{r})$ is centered at $\bm{r}_\alpha$, i.e., $\phi_\alpha(\bm{r+r_\alpha})$ appearing in the integral is centered around the coordinate system's origin, and the phase factor $\mathrm{e}^{-i\bm{p\cdot r}_\alpha}$ encodes the position of the orbital within the unit cell. Summing the Bloch phases $\mathrm{e}^{i\bm{k\cdot R}}$ over all lattice vectors $\bm{R}$ enforces in-plane momentum conservation, $\bm{p}_{\parallel} = \bm{k}_{\parallel}$, where we restrict $\bm{k}_\parallel$ to the first Brillouin zone; the broken translational symmetry normal to the surface causes the normal component $p_\perp$ to be photon energy dependent (see Sec.~\ref{Sec:SM_kz_determination}). Energy conservation relates the binding energy $\omega$, the photon energy $E$, and the photoelectron kinetic energy $E_\mathrm{kin}$. The attenuation length $\lambda$ is introduced phenomenologically as an effective combination of the penetration depth of the incident photon field and the escape depth of the outgoing photoelectron. With the surface at $z=0$ and the solid occupying $z<0$, the factor $\mathrm{e}^{z_\alpha/\lambda}$ attenuates contributions from orbitals deeper inside the sample. Note that the ACA directly causes a breaking of gauge invariance of the electromagnetic field, meaning that, e.g., velocity gauge would lead to different results \cite{Yen2024}. The integral in Eq.~(\ref{eq:new_M_wannier}) vanishes for certain orbital symmetries, indicating regions in the ARPES map with zero intensity. For atomic-like $d$-orbitals, as well as hybridized Wannier functions, this can be observed in Sec.~\ref{Sec:SM_matrix_elements}.

The ARPES intensity is obtained by coherently summing the amplitudes of all Wannier orbitals,
\begin{align}
    I(\bm{k}, E) \sim \sum_{\alpha,\beta} {M}^*_\alpha(\bm{k},E) \; {M}_\beta(\bm{k},E) \; {A}_{\alpha \beta}(\bm{k},\omega), \label{eq:new_ARPES_master}
\end{align}
which is Eq.~(\ref{eq:ARPES_signal_main}) of the main text and the only expression used in our simulations. The weights are given by the DMFT lattice spectral function in the Wannier basis,
\begin{align}
    \bm{A}(\bm{k}, \omega) &= \frac{-1}{2\pi i} \Big[ \bm{G}(\bm{k}, \omega) - \bm{G}^\dagger(\bm{k}, \omega) \Big], \label{eq:new_A_def}\\
    \bm{G}(\bm{k}, \omega) &= \Big[ (\omega+\mu)\bm{1} - \bm{H}^{\mathrm{DFT}}(\bm{k}) - \bm{\Sigma}(\omega) \Big]^{-1}, \label{eq:new_dyson}
\end{align}
where $\mu$ is the chemical potential, the double-counting correction is absorbed into $\bm{\Sigma}(\omega)$, and we stress that the imaginary part is not taken element-wise but defined via the Hermitian conjugate. The Dyson equation (\ref{eq:new_dyson}) is naturally formulated in the Wannier basis, since the DMFT self-energy $\bm{\Sigma}(\omega)$ is local: it is $\bm{k}$ independent, diagonal in the orbital index of the correlated Ni-$3d$ and Nd-$5d$ Wannier functions, and vanishes on the uncorrelated interstitial-$s$ orbital. In contrast, $\bm{H}^{\mathrm{DFT}}(\bm{k})$ contains $\bm{k}$-dependent hopping between different Wannier orbitals, so that $\bm{G}$ and $\bm{A}$ are in general not diagonal in $\alpha,\beta$. The off-diagonal terms $\alpha\neq\beta$ in Eq.~(\ref{eq:new_ARPES_master}) describe the interference of photoemission amplitudes from different orbitals; retaining only the diagonal terms yields the weighted-trace approximation discussed in the main text.

To simulate a realistic signal, including contributions from the whole crystal, one would need to explicitly construct a slab and consider the interference of different layers, while taking into account the finite penetration depth. Here we take a simplified route: Assuming a 2D system with $N$ equivalent layers, separated by the lattice constant $c$, and further taking infinite penetration length $\lambda\rightarrow\infty$, we see from Eq.~(\ref{eq:new_M_wannier}) that the matrix elements are mainly determined by the shape of the Wannier orbitals, while each layer $l$ picks up a phase factor $\mathrm{e}^{ip_zc\,l}$ ($c$: $c$-axis lattice constant). The sum over all orbitals in Eq.~(\ref{eq:new_ARPES_master}) then decomposes into a sum over layers $l,l'$ and a sum over orbitals $\alpha,\beta$ within each layer. The sum over layers takes the form of a Fourier transform. The layer-dependent spectral function $A_{\alpha\beta}$ (which formally depends on the 2D momentum $\bm{k}$) then becomes the bulk spectral function in the limit $N\rightarrow\infty$, and the layer dependence translates into a dependence on the quantum number $k_z$:
\begin{align}
    I(\bm{k}, E) \sim \sum_{\alpha,\beta} {M}^*_\alpha(\bm{k},E) \; {M}_\beta(\bm{k},E) \; {A}_{\alpha \beta}(\bm{k},k_z,\omega). \label{eq:new_ARPES_signal_kz}
\end{align}
This entirely neglects surface effects on the ARPES photocurrent, which are expected to become increasingly relevant at lower photon energies.

\section{Determination of \texorpdfstring{$k_z$}{kz}}
\label{Sec:SM_kz_determination}

\begin{figure}[tb]
    \centering
    \includegraphics[width=0.5\textwidth]{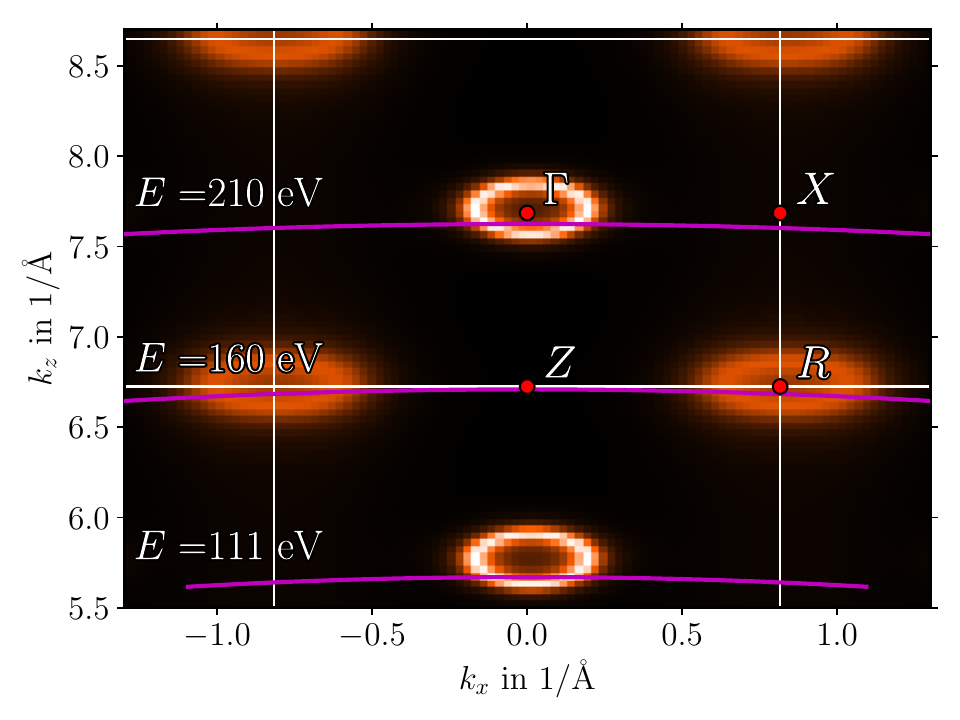}
    \caption{\label{fig:kz} Value of $k_z$ for different photon energies $E$ within $\bm{k}$-space in magenta. The false color shows the  DMFT Fermi surface in the $k_y=0$-plane, with pronounced pockets around both the $\Gamma$- and $R$-point. White lines are Brillouin zone boundaries.}
\end{figure}

In a three-dimensional bulk crystal, $k_z$ labels the initial electronic states. At the surface, however, translational symmetry is broken along the surface-normal direction, and $k_z$ is therefore not conserved when the photoelectron enters the vacuum. Its value must instead be inferred from the measured kinetic energy and emission angle using a model for the photoelectron final state. We assume a free-electron final state.

For a photon energy $E$, energy conservation gives the photoelectron kinetic energy in vacuum as
\begin{align}
    E_\mathrm{kin}=E-E_b-\Phi,
\end{align}
where $E_b\geq 0$ is the binding energy relative to the Fermi level and $\Phi$ is the work function. We denote the photoelectron momentum in vacuum by $\bm{p}$. Its components parallel and perpendicular to the surface are
\begin{align}
    p_\parallel & = {\sqrt{2mE_\mathrm{kin}}}\sin\theta,\\
    p_\perp & = {\sqrt{2mE_\mathrm{kin}}}\cos\theta,
\end{align}
where $\theta$ is the emission angle measured from the surface normal. The potential step at the surface preserves the parallel component of the final-state wave vector, whereas its perpendicular component changes. Introducing the inner potential $V_0$, conservation of energy gives 
\begin{align}
    E_\mathrm{kin} = \frac{\hbar^2 (k_\parallel^2+k_\perp^2)}{2m} - V_0 = \frac{p_\parallel ^2 + p_\perp ^2}{2m}  \; ,   
\end{align}
and the free-electron final-state wave vector inside the crystal is therefore
\begin{align}
    k_{f,\parallel} & = p_\parallel/\hbar,\\
    k_{f,z} & = \frac{1}{\hbar}\sqrt{2m\left(E_\mathrm{kin}\cos^2\theta+V_0\right)}.
    \label{eq:new_kfz}
\end{align}

Neglecting the photon momentum, crystal-momentum conservation relates the initial- and final-state wave vectors according to $\bm{k}_i=\bm{k}_f+\bm{G}$, where $\bm{G}$ is a reciprocal-lattice vector. 
At the Fermi level, $E_b=0$. Using $E=160$\,eV, $\Phi=5.5$\,eV, and $V_0=17$\,eV as used by Li et al.~\cite{li2025}, Eq.~(\ref{eq:new_kfz}) gives $k_{f,z}\simeq 7\pi/c$. In the reduced-zone description,
this corresponds to the initial-state plane $k_{i,z}\simeq\pi/c$, i.e., the $Z$--$R$ plane. 

At fixed photon energy, Eq.~(\ref{eq:new_kfz}) also shows that $k_{f,z}$ decreases with increasing emission angle. A constant-photon-energy Fermi surface map therefore follows a curved trajectory in $(k_\parallel,k_z)$ space, as illustrated in Fig.~\ref{fig:kz}, rather than an exactly constant-$k_z$ plane.
However, due to the minor deviation at $R$-momentum for $E = 160$\,eV, we approximate the probed $k_z$ as constant.

The resulting $k_z$ assignment remains model dependent and should be understood as an approximate assignment within the free-electron final-state model.

\newpage

\section{Matrix elements for different orbitals}
\label{Sec:SM_matrix_elements}
\subsection{Wannier matrix elements}
\label{Sec:SM_wannier_matrix_elements}

\begin{figure}[tb]
    \centering
    \includegraphics[width=\textwidth]{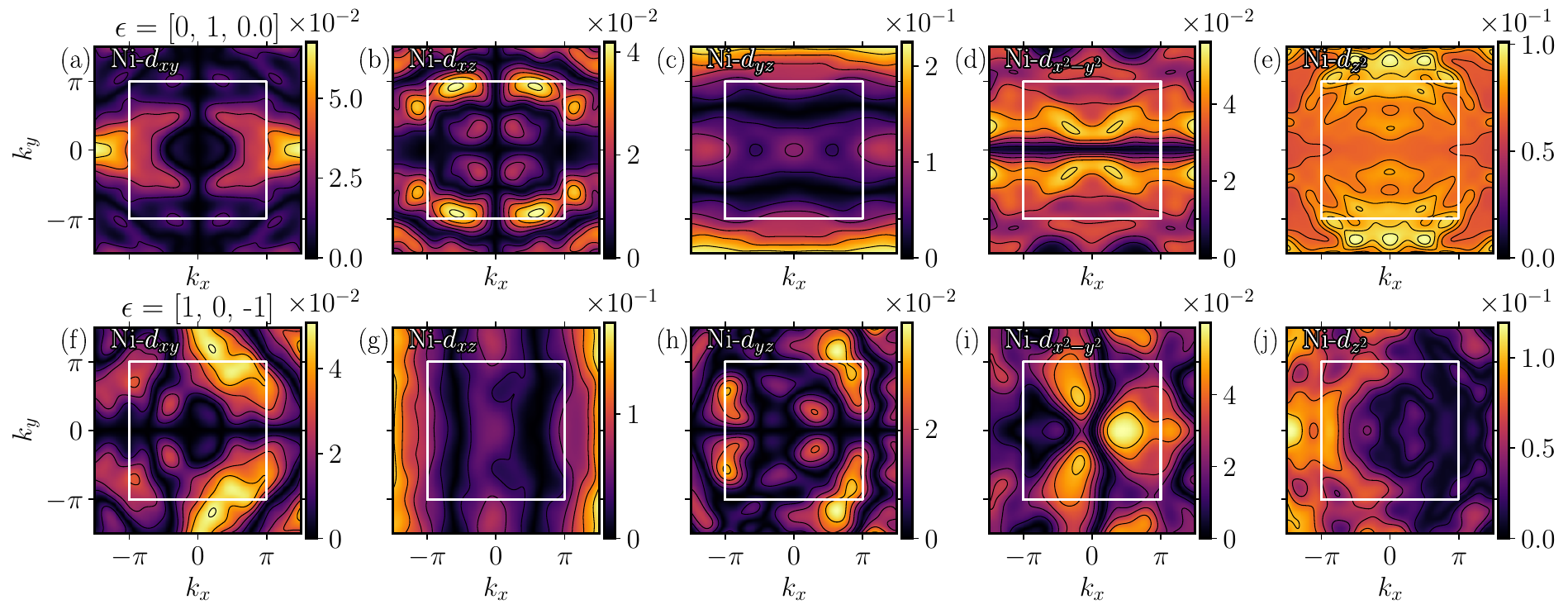}
    \caption{\label{fig:M_Ni} Absolute values of matrix elements for Ni-derived Wannier orbitals in LV (upper row) and LH (lower row).}
\end{figure}
\begin{figure}[tb]
    \centering
    \includegraphics[width=\textwidth]{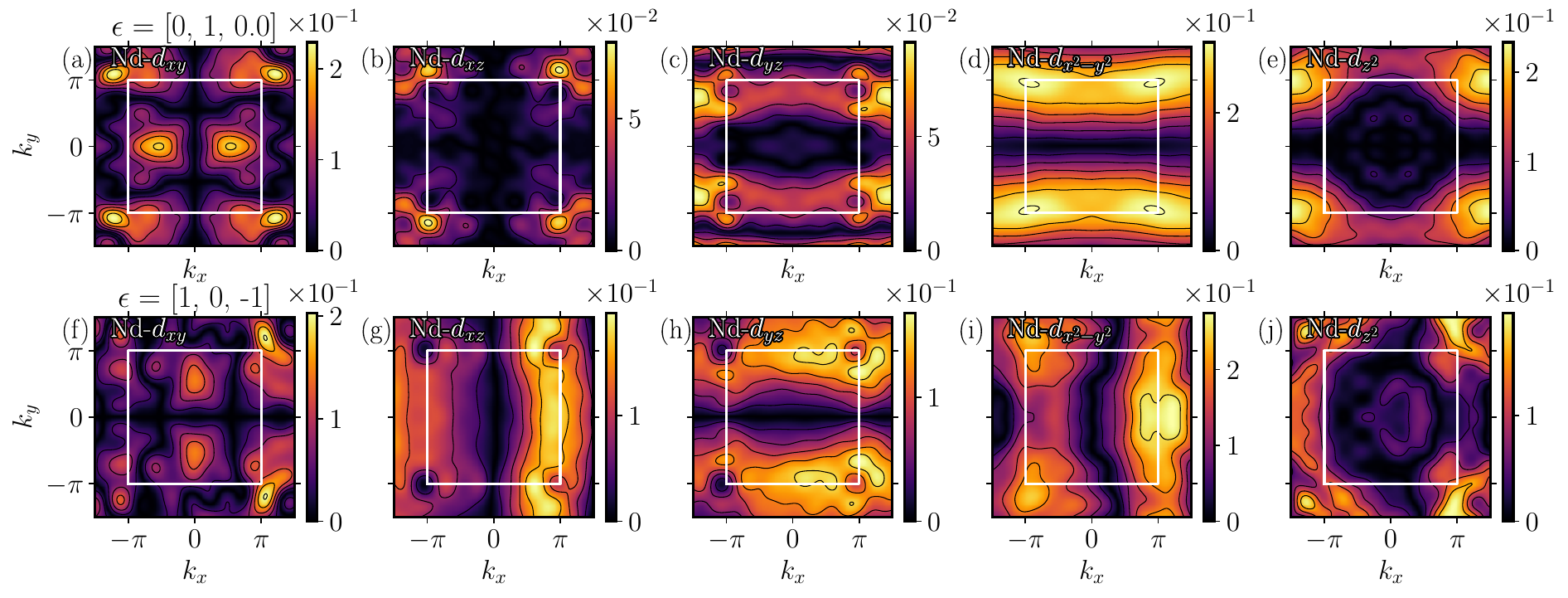}
    \caption{\label{fig:M_Nd} Absolute values of matrix elements for Nd-derived Wannier orbitals in LV (upper row) and LH (lower row).}
\end{figure}
\begin{figure}[tb]
    \centering
    \includegraphics[width=1\textwidth]{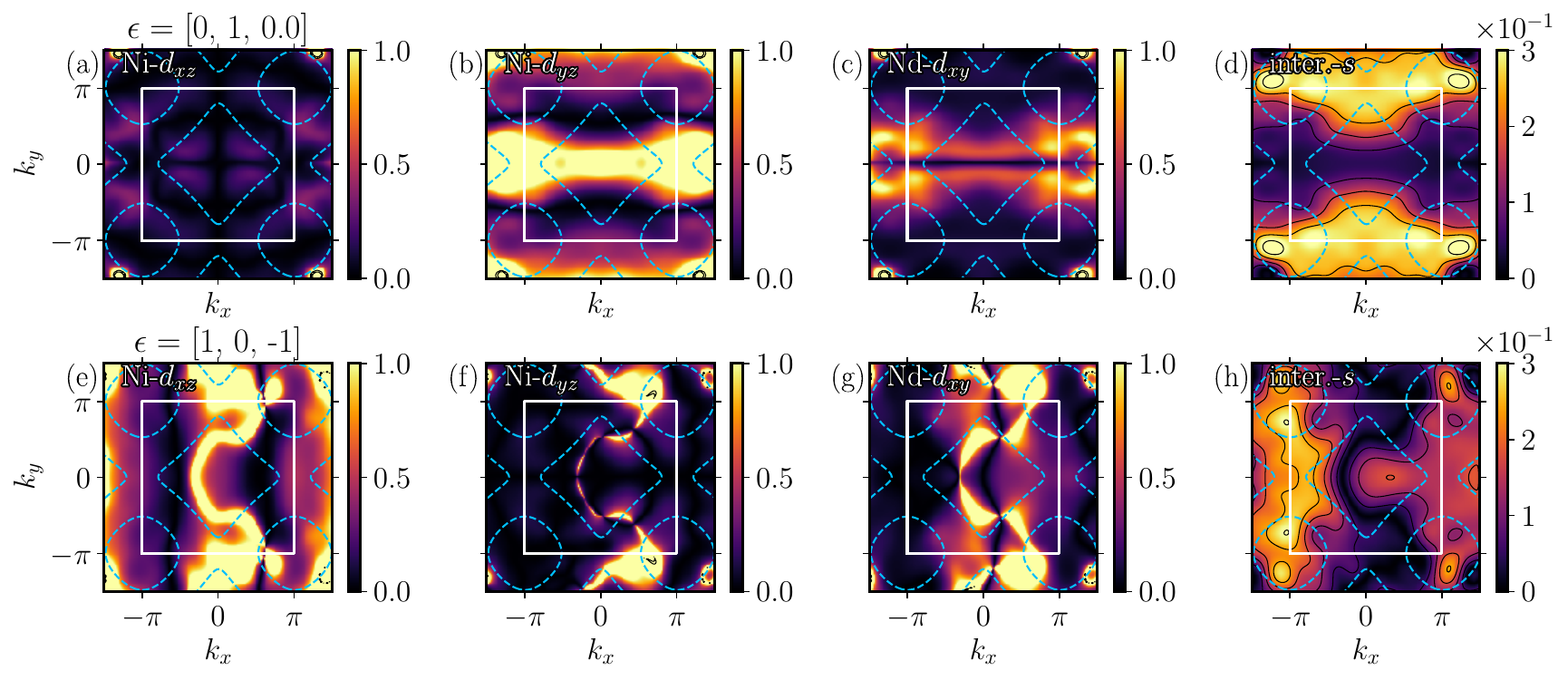}
    \caption{\label{fig:M_s} (a--c, e--g) Ratios between absolute values of the matrix elements for different orbitals to interstitial-$s$ orbital in LV (a)-(c) and LH (e)-(g). The colorscale is cropped at a ratio of 1, to highlight that relevant regions in $\bm{k}$-space show a ratio of less than 1. (d,h) Absolute values of matrix elements for the interstitial-$s$ orbital in LV and LH polarization, respectively. Blue-dashed lines indicate the DFT Fermi contour.}
\end{figure}

\begin{figure}[tb]
    \centering
    \includegraphics[width=\textwidth]{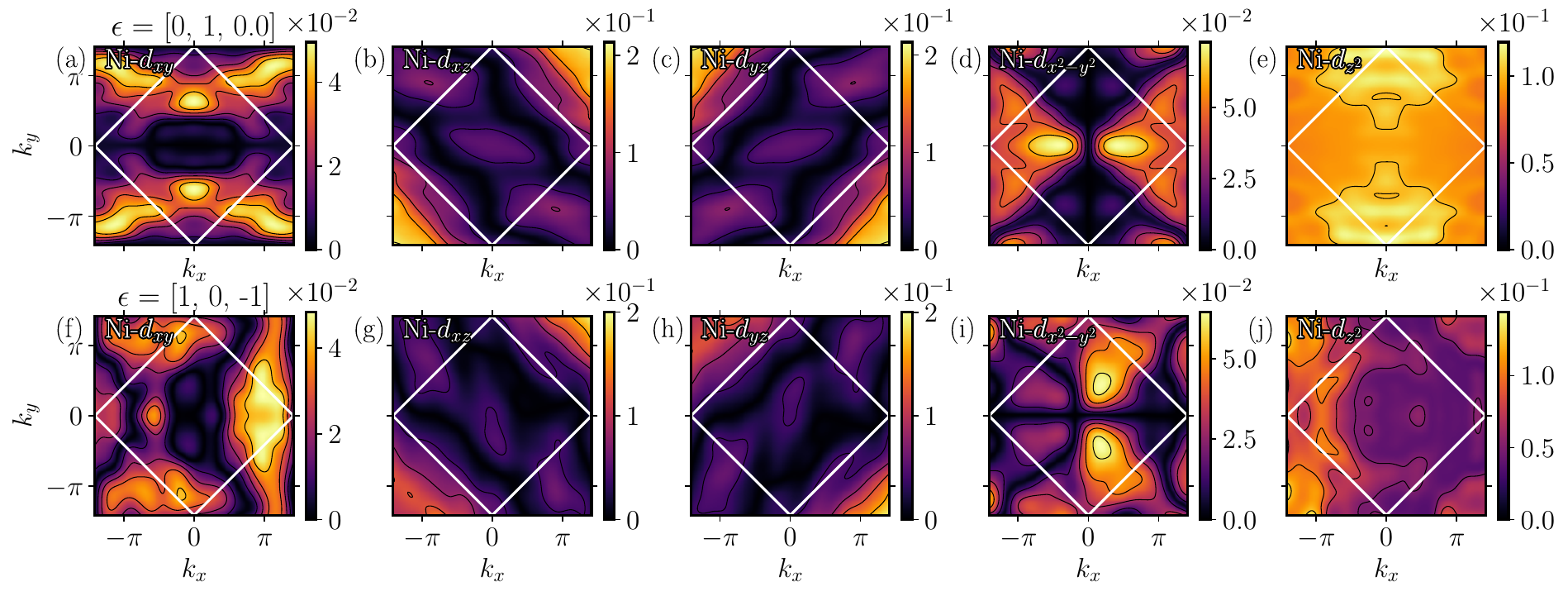}
    \caption{\label{fig:M_rot_Ni} Absolute values of matrix elements for Ni-derived Wannier orbitals in LV (upper row) and LH (lower row), calculated in rotated basis.}
\end{figure}
\begin{figure}[tb]
    \centering
    \includegraphics[width=\textwidth]{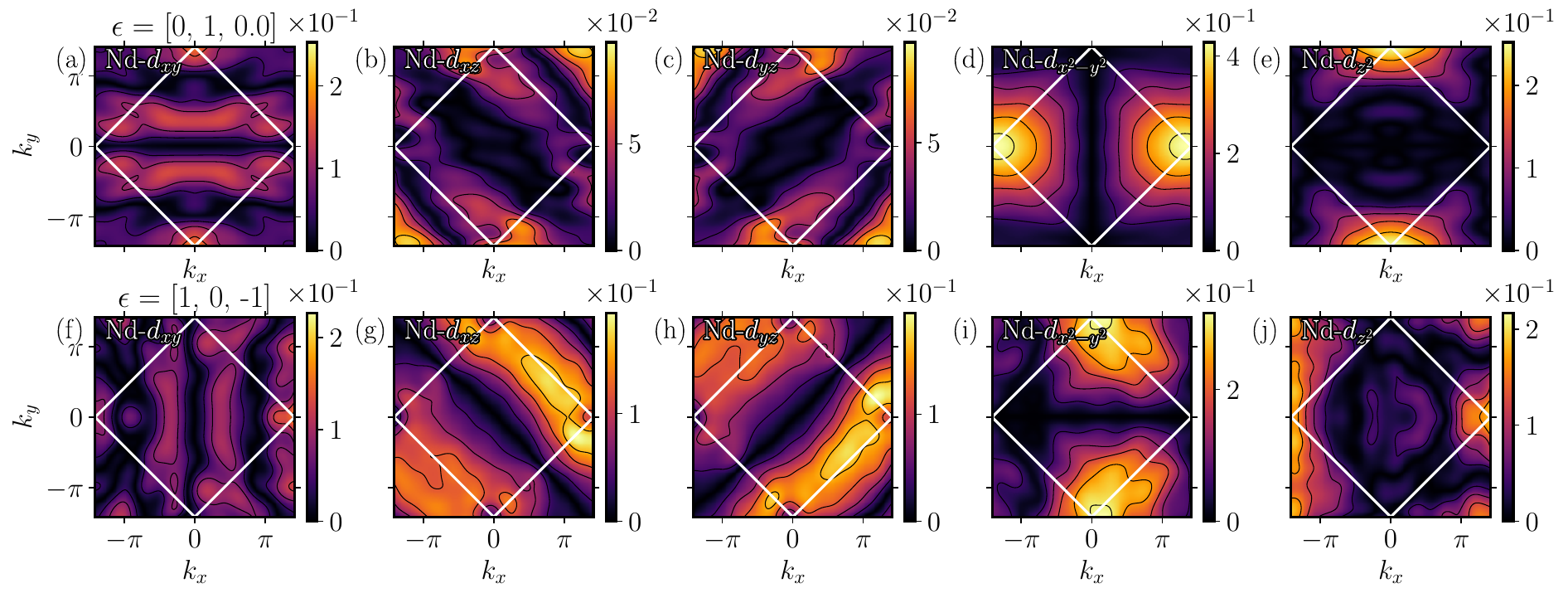}
    \caption{\label{fig:M_rot_Nd} Absolute values of matrix elements for Nd-derived Wannier orbitals in LV (upper row) and LH (lower row), calculated in rotated basis.}
\end{figure}
\begin{figure}[tb]
    \centering
    \includegraphics[width=1\textwidth]{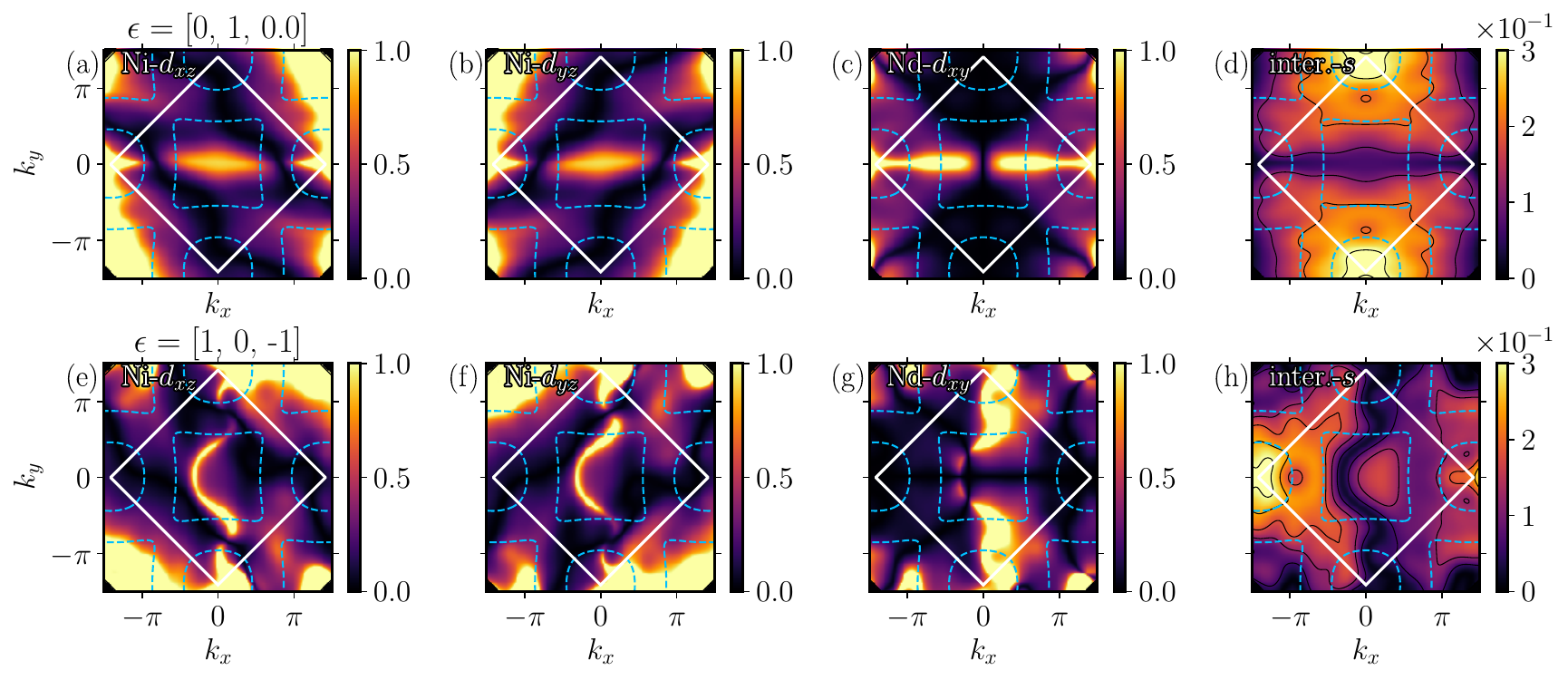}
    \caption{\label{fig:M_rot_s} Same as Fig.~\ref{fig:M_s}, but calculated in rotated basis.}
\end{figure}
Figs.~\ref{fig:M_Ni}-\ref{fig:M_rot_s} show the absolute value of our calculated matrix elements for all 11 orbitals in the local Wannier basis. Note that by symmetry, matrix elements for $d$-like (i.e., inversion symmetric), origin-centered orbitals are purely imaginary. However, not all Wannier orbitals are localized at the same atomic site, which leads to relative phase-factors causing interference patterns between realistic Wannier orbitals \cite{dtg3-66t9,Yen2024}.\\
In the first three rows of Fig.~\ref{fig:M_s} and \ref{fig:M_rot_s}, we additionally show the absolute value of Ni-$d_{xz}$, Ni-$d_{yz}$, and Nd-$d_{xy}$ matrix elements, each divided by the absolute value of the interstitial-$s$ matrix element. The ratios illustrate what we advocate in the main text: The regions in $k$-space, where the $A$-pocket is bright in ARPES is dominated by the interstitial-$s$ matrix elements.\\
Let us note that this effect lies beyond the modeling presented in Ref.~\cite{Ye_2013} and used in Ref.~\cite{li2025} which relies on pure atomic $d$-orbitals, as also shown in Fig.~\ref{fig:ME_examples} and Fig.~\ref{fig:ME_examples_chinook}.

\subsection{Comparison atomic-like orbitals}
\label{Sec:SM_atomic_matrix_elements}
\begin{figure}[tb]
    \centering
    \includegraphics[width=\textwidth]{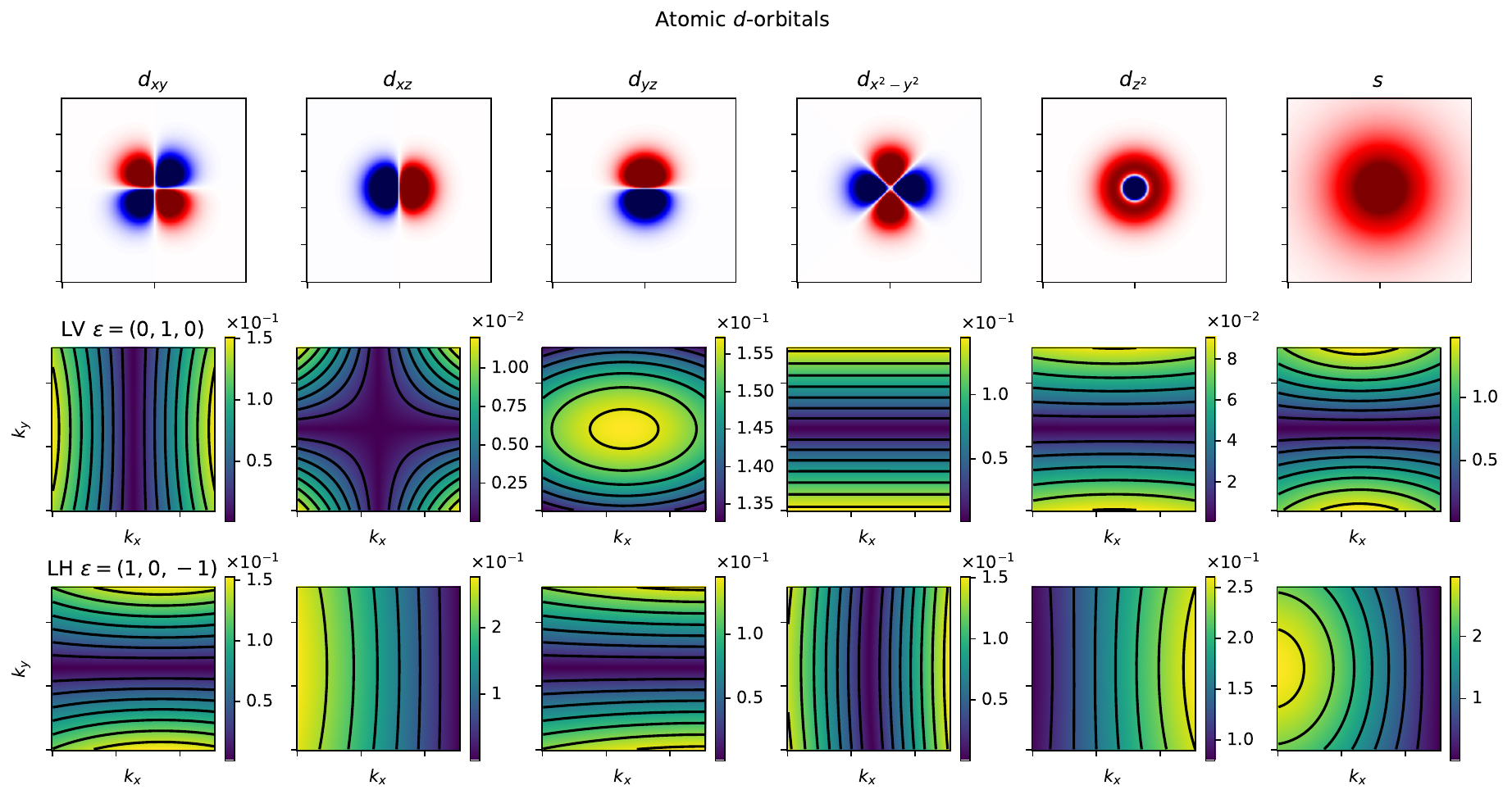}
    \caption{\label{fig:ME_examples} Matrix elements for atomic-like $d$-orbitals calculated for LV (center row) and LH (lower row). The upper row plots the respective charge density (colored by the wavefunction sign).}
\end{figure}
As a comparison to the matrix elements, corresponding to Wannier functions, as shown above, we also compute matrix elements for atomic-like orbitals. They are presented in Fig.~\ref{fig:ME_examples}. They are similar to the ones calculated by Ye et al. in Ref.~\cite{Ye_2013}. The overall symmetries of atomic-like $d$-orbital matrix elements can generally be found in the Wannier orbital matrix elements as well, where the latter become more complex due to hybridizations with, e.g., O-$p$ orbitals.

\subsection{Comparison to chinook}
\begin{figure}[tb]
    \centering
    \includegraphics[width=\textwidth]{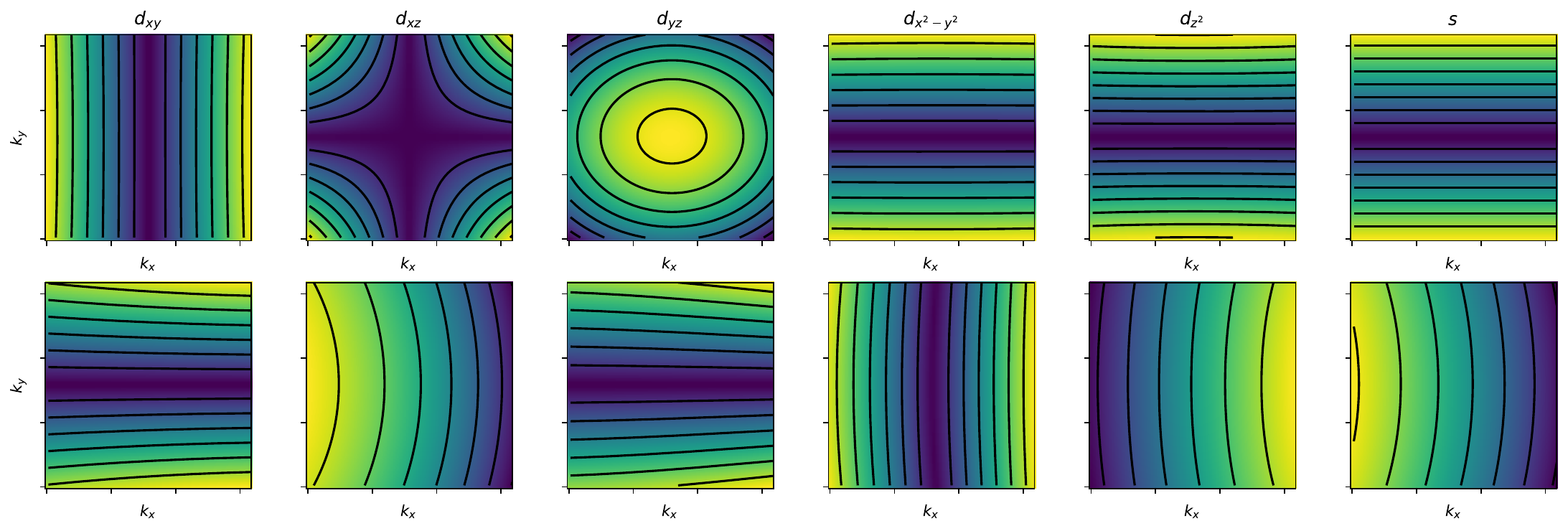}
    \caption{\label{fig:ME_examples_chinook} Matrix elements for atomic-like $d$-orbitals in LV (center row) and LH (lower row) polarization, calculated with the \textsc{chinook} package \cite{Day2019}.}
\end{figure}

To validate our implementation for calculating matrix elements, we additionally computed the matrix elements of atomic-like $d$-orbitals and an interstitial-$s$ orbital using the independent, open-source package \textsc{chinook} \cite{Day2019}. As shown in Fig.~\ref{fig:ME_examples_chinook}, this reproduces our results qualitatively; in particular, it correctly captures the opposite response of the $d_{z^2}$ and interstitial-$s$ orbitals to the out-of-plane ($z$) component of the polarization.

\subsection{Hybridized orbitals}
\label{Sec:SM_hybridized_orbitals}

\begin{figure}[tb]
    \centering
    \includegraphics[width=0.6\textwidth]{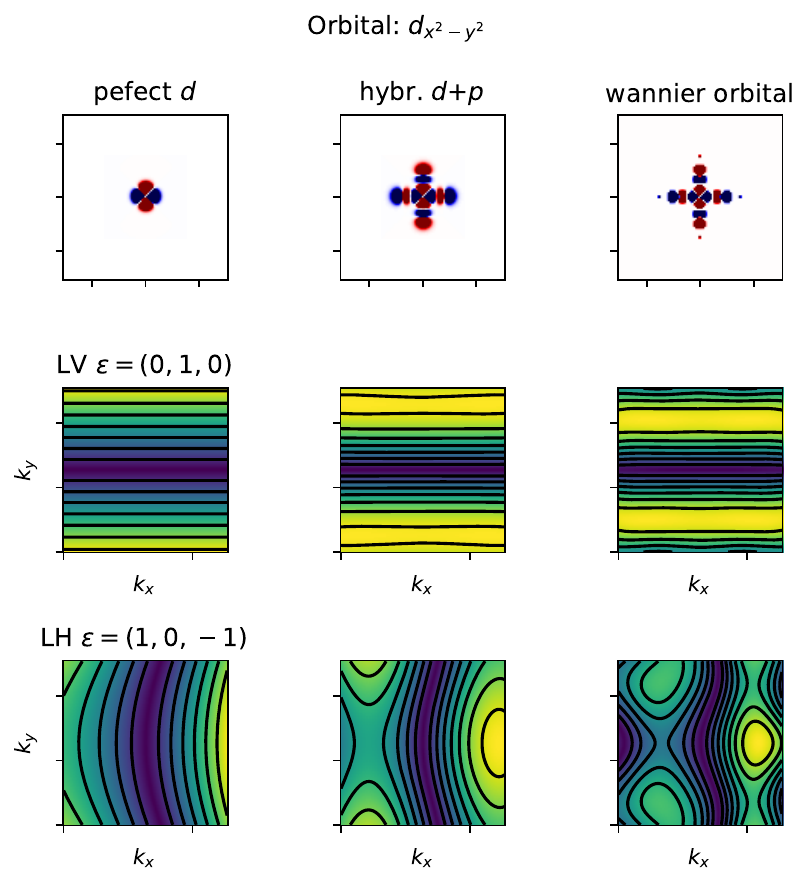}
    \caption{\label{fig:ME_example_dx2y2} Matrix elements for $d_{x^2-y^2}$-derived orbitals. Left column: Atomic-like $3d_{x^2-y^2}$ orbital; Center column: Hybridized orbital, consisting of an atomic-like $3d_{x^2-y^2}$ orbital and 4 neighboring, atomic-like O-$2p_{x/y}$ orbitals; Right column: Maximally localized Wannier function for NdNiO$_2$. The upper row plots the charge density (colored by the wavefunction sign), the center (lower) row shows the absolute value of the corresponding matrix element in LV (LH) polarization.
    }
\end{figure}

To illustrate the effect of oxygen hybridization, Fig.~\ref{fig:ME_example_dx2y2} compares a pure atomic-like $3d_{x^2-y^2}$ orbital, a linear combination of this orbital with neighboring O-$2p$ orbitals located a distance $\pm a/2$ in $x$- and $y$-direction from the central site, and the corresponding maximally localized Wannier orbital of NdNiO$_2$. 
We construct the hybridized orbital $\phi(\bm{r})$ as
\begin{align}
    \phi(\bm{r}) = \frac{1}{\sqrt{\alpha^2+4}}  \bigg[
    \alpha\,&\Psi_{d_{x^2-y^2}}(\bm{r}) \\
    +& \Psi_{p_x}(\bm{r}-\frac{a}{2}\hat{x}) -\Psi_{p_x}(\bm{r}+\frac{a}{2}\hat{x})  \\
    +& \Psi_{p_y}(\bm{r}+\frac{a}{2}\hat{y}) -\Psi_{p_y}(\bm{r}-\frac{a}{2}\hat{y})
     \bigg].
\end{align}
Here, $\Psi_{d_{x^2-y^2}}$, $\Psi_{p_x}$, and $\Psi_{p_y}$ corresponds to an atomic $d_{x^2-y^2}$, $p_x$, and $p_y$ orbital, respectively. The lattice constant is $a$, and $\hat{x}$ ($\hat{y}$) is the unit vector in the $x$-($y$-)direction. We heuristically choose a hybridization parameter $1/\alpha = 0.06$, to match the shape of Wannier orbital and its matrix element.\\
The Wannier matrix elements deviate from those of the pure $d_{x^2-y^2}$ orbital, but their main features are reproduced by including the neighboring O-$p$ contributions in the linear combination. 
This example illustrates that oxygen hybridization accounts for the principal deviations from the atomic-like matrix elements, and can also be observed for the other $d$-orbitals.

\newpage

\section{Setting \texorpdfstring{$d$}{d}-channels dark}
\label{Sec:SM_channels_dark}
\begin{figure}[tbp]
    \centering
    \includegraphics[width=0.44\textwidth]{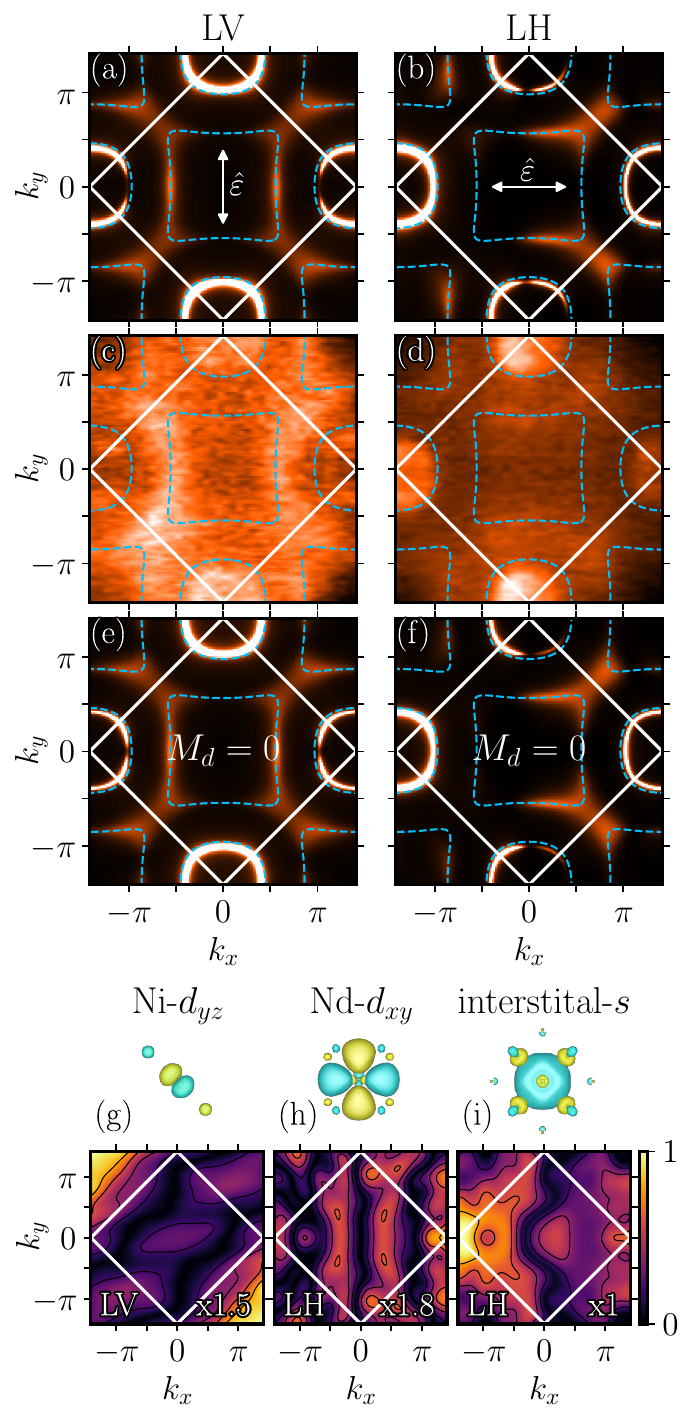}\hfill
    \includegraphics[width=0.44\textwidth]{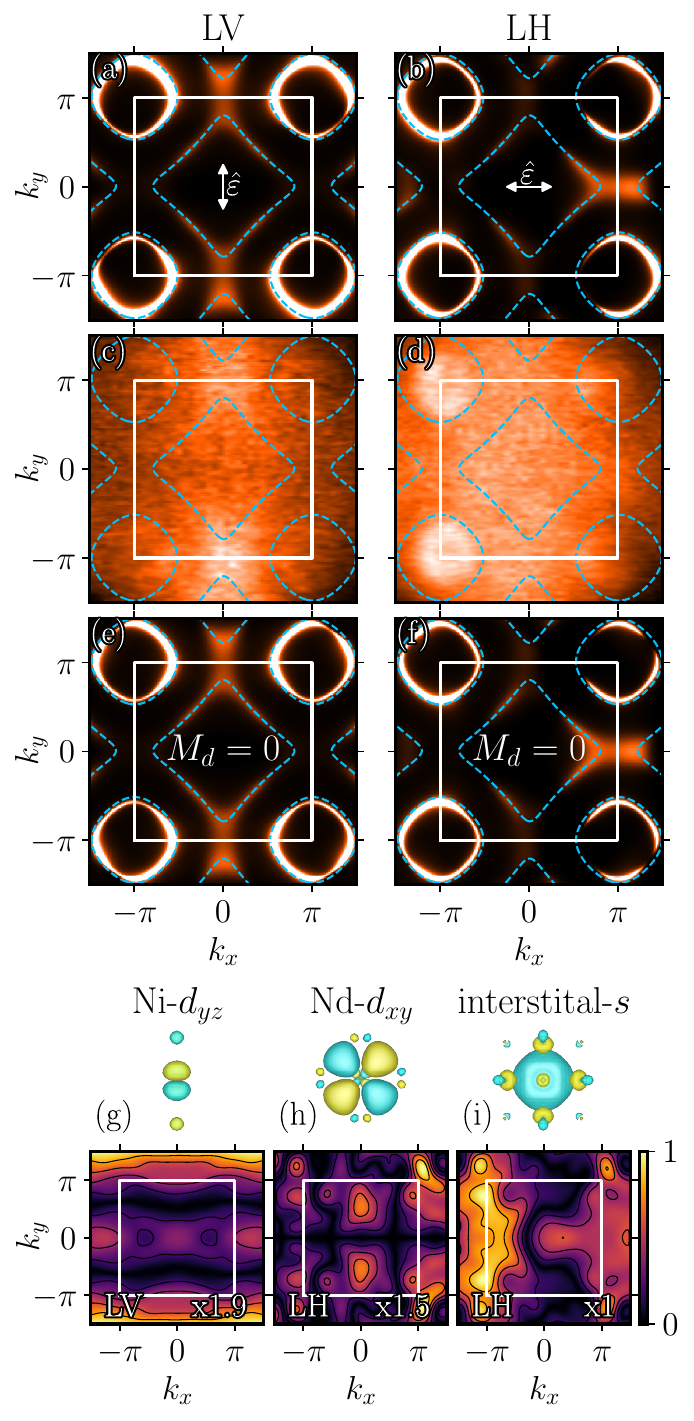}
    \caption{\label{fig:SM_no_d} Same as Figs.~\ref{fig:ARPES_rot} (left, rotated basis) and \ref{fig:ARPES} (right, conventional basis) of the main text, but with the Ni-$d_{xz/yz}$ and Nd-$d_{xy}$ matrix elements set to zero in the third row, $M_d\equiv 0$ [panels (e),(f)], instead of $M_s\equiv 0$.}
\end{figure}

In the main text, we present results for setting $M_s$ dark. Now we also do the opposite and set $M_d$ dark for the other possible contributions to the $A$-pocket.
Figure~\ref{fig:SM_no_d} repeats the simulations of Figs.~\ref{fig:ARPES_rot} and \ref{fig:ARPES} of the main text, but in the third row the three $d$-derived channels contributing to the $A$-pocket are switched off simultaneously, $M_{d_{xz/yz}}\equiv M_{d_{xy}}\equiv 0$, with everything else unchanged. In contrast to $M_s\equiv 0$, the $A$-pocket intensities in panels (e) and (f) are almost indistinguishable from the full calculation in panels (a) and (b), in both experimental setups. For the present photon energy and geometry, the $d$-derived matrix elements are thus too small near the $A$-pocket to leave a visible imprint on the maps [cf.\ the matrix-element ratios in Figs.~\ref{fig:M_s} and \ref{fig:M_rot_s}], even though these orbitals carry about two thirds of the spectral weight of the pocket (Sec.~\ref{Sec:results_DMFT} of the main text).

\section{Experimental setup}
\label{Sec:SM_experimental_setup}
\begin{figure}[tb]
    \centering
    \includegraphics[width=0.4\textwidth]{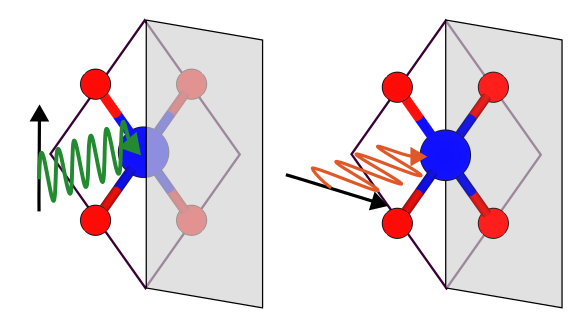}
    \caption{\label{fig:arpes_sketch} Experimental geometry of ARPES measurements in rotated basis, employed in \cite{li2025}. Left: LV, right: LH polarization.}
\end{figure}
Figure~\ref{fig:arpes_sketch} shows the experimental setup of ARPES measurements performed in the 45° rotated basis, as employed by Li et al. \cite{li2025} and reconstructed in our simulations.
While LV (left side; green) is uniquely defined as $\hat{\varepsilon}_{\mathrm{LV}}=(0,1,0)$, LH polarization (orange; right side) has an out-of-plane component which depends on the photon incidence angle. We choose $\hat{\varepsilon}_{\mathrm{LH}}=(1,0,-1)$, which corresponds to an incidence angle of 45°.

\section{Comparison to \texorpdfstring{LaNiO$_2$}{LaNiO2}}
\label{Sec:SM_lanio2_comparison}
\begin{figure}[tb]
    \centering
    \includegraphics[width=0.8\textwidth]{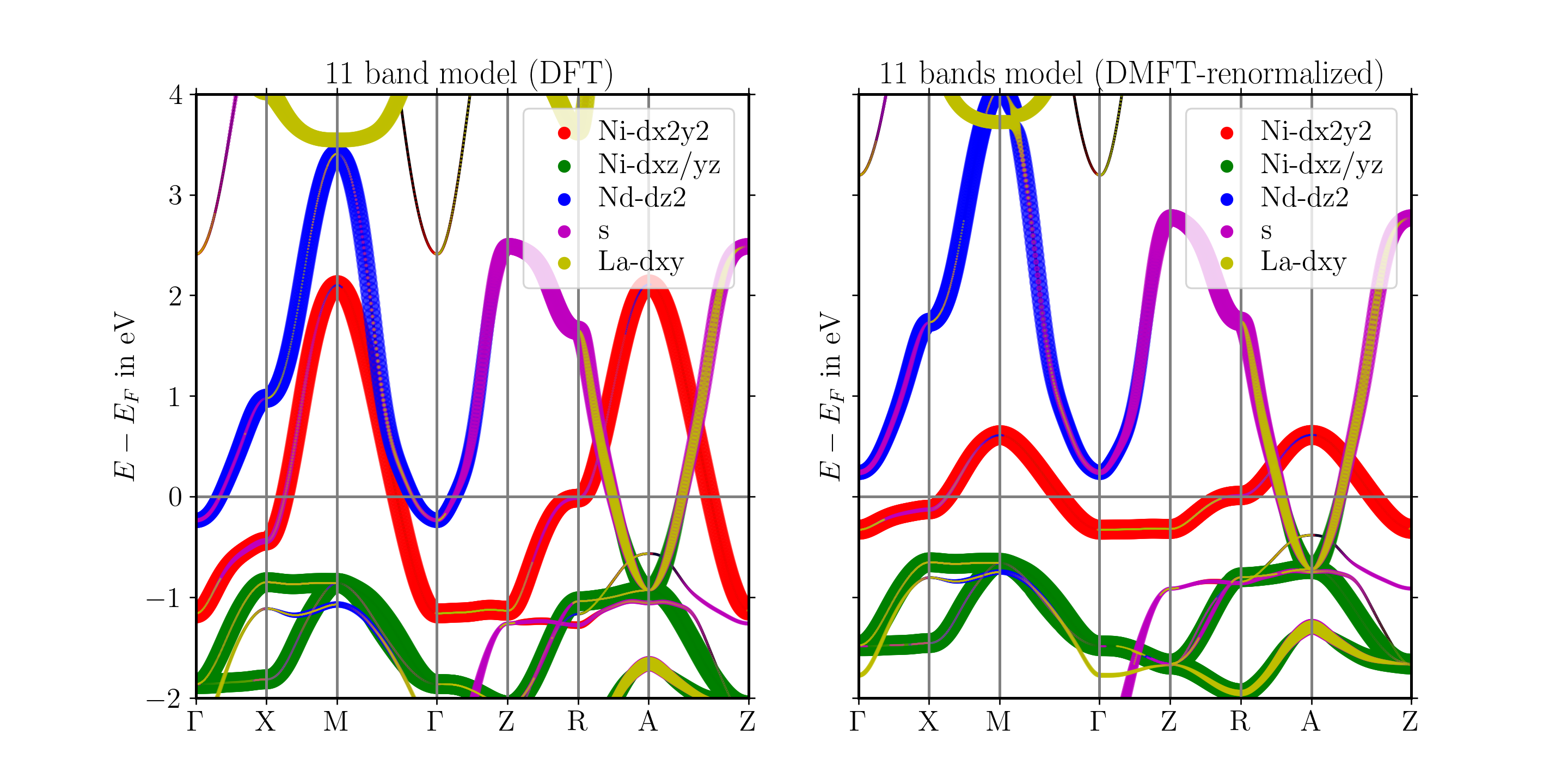}
    \caption{\label{fig:SM_bands_sketch_LNO} Band structure of LaNiO$_2$. Left shows DFT bands; right is DMFT-renormalized band structure.}
\end{figure}
\begin{figure}[tb]
    \centering
    \includegraphics[width=0.6\textwidth]{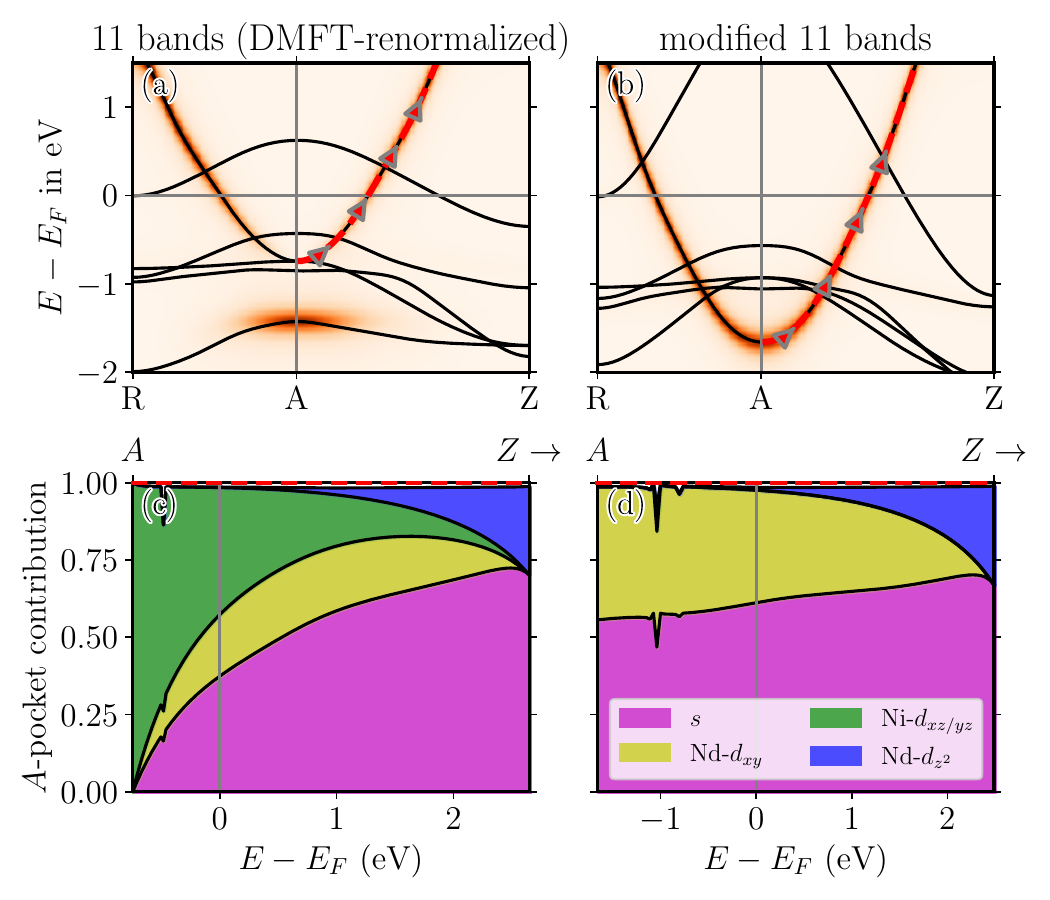}
    \caption{\label{fig:SM_band_pros_LNO} (a),(b) Band structure around the $A$-pocket in bulk LaNiO$_2$ for the physical, DMFT-renormalized 11-band model (a) and an artificial model without hybridization of La-$d_{xy}$ and interstitial-$s$ to Ni-$d_{xz/yz}$ (b), analogous to Fig.~\ref{fig:DFT_pros} in the main text. The colormap indicates interstitial-$s$ spectral weight; the tracked $A$-pocket band is highlighted by the red dashed line and gray arrows. (c),(d) Orbital contribution to the $A$-pocket as a function of energy along $A$--$Z$, for the models of (a) and (b), respectively.}
\end{figure}
Since LaNiO$_2$ is electronically very similar to NdNiO$_2$, we repeat our DFT+DMFT and ARPES simulations for this related compound.

In LaNiO$_2$, the local DMFT correlations shift the $\Gamma$-pocket entirely above the Fermi energy, in contrast to NdNiO$_2$, where a small $\Gamma$-pocket survives in our DMFT calculation despite being absent from experiment (see main text). This can be seen by comparing the DFT and DMFT-renormalized band structures in Fig.~\ref{fig:SM_bands_sketch_LNO}. 

As in the main text, Fig.~\ref{fig:SM_band_pros_LNO} shows the orbital character of the $A$-pocket as a function of energy. At the DMFT (DFT) Fermi level, we find contributions of 38(39)\%, 20(22)\%, and 41(38)\% from interstitial-$s$, La-$d_{xy}$, and Ni-$d_{xz/yz}$ states, respectively---a slight increase in interstitial-$s$ character compared to NdNiO$_2$, but otherwise a qualitatively similar, strongly hybridized $A$-pocket. This confirms that the hybridized character of the $A$-pocket found in the main text is a generic feature of infinite-layer nickelates. Note that the bump seen in Fig.~\ref{fig:SM_band_pros_LNO} (d) around $-1$\,eV is a numerical artifact.
%not an artifact of the (experimentally absent) $\Gamma$-pocket present in our NdNiO$_2$ calculation.

\begin{figure}[tb]
    \centering
    \includegraphics[width=0.5\textwidth]{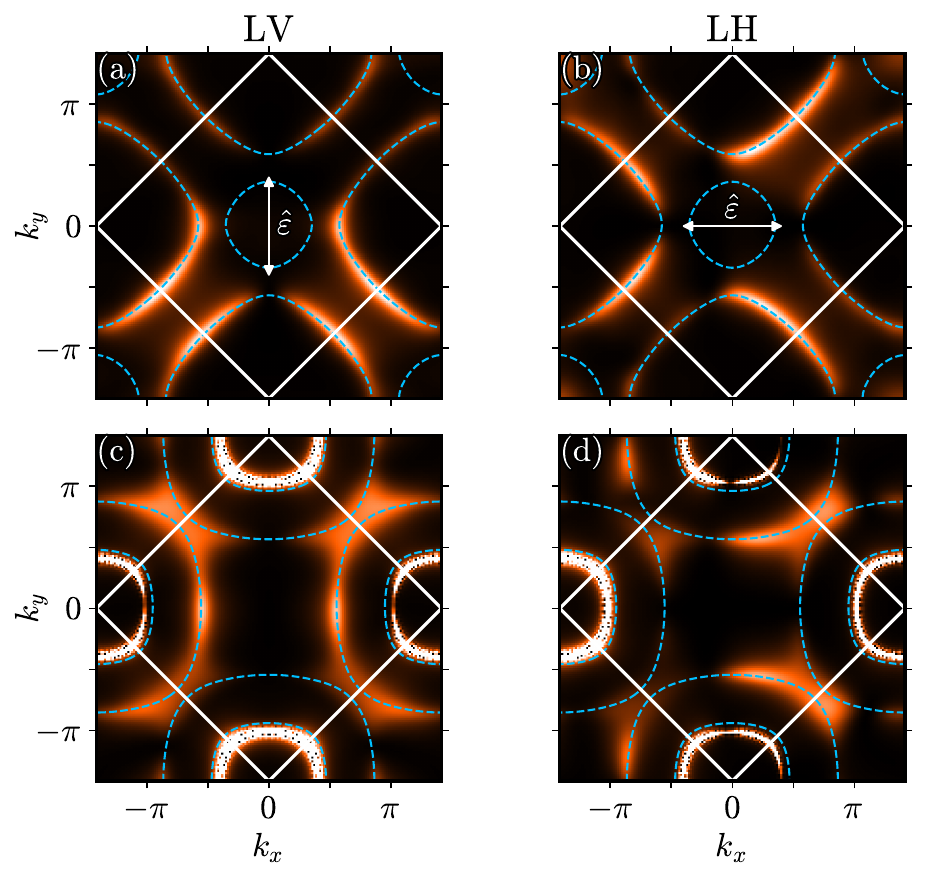}
    \caption{\label{fig:ARPES_LNO} Simulated Fermi surface of LaNiO$_2$ for (a,b) $k_z=0$ and (c,d) $k_z=\pi$. White squares indicate the first Brillouin zone; blue dashed lines are the DFT Fermi contour.}
\end{figure}

Fig.~\ref{fig:ARPES_LNO} shows our simulated ARPES Fermi surface of LaNiO$_2$ for LV and LH polarization, both in the $k_z=0$ and $k_z=\pi$ plane.
\end{document}